\documentclass[aps, prd, twocolumn, nofootinbib,]{revtex4-2}

\usepackage{amsmath}
\usepackage{amsfonts}
\usepackage{wasysym}
\usepackage{graphicx}
\usepackage{comment}
\usepackage{tensor}

\def\filetype{pdf}

\def\path{}

\allowdisplaybreaks[1]

\begin{document}

\title{Dynamical evolution of dark matter admixed neutron stars}
\author{Troy Gleason, Ben Brown, and Ben Kain}
\affiliation{Department of Physics, College of the Holy Cross, Worcester, Massachusetts 01610, USA}

\begin{abstract}
\noindent We dynamically evolve for the first time dark matter admixed neutron stars with fermionic dark matter.  These systems are mixtures of the ordinary nuclear matter of a neutron star and dark matter.  To perform our dynamical evolutions, we derive the equations of motion, in conservation form, for spherically symmetric systems with an arbitrary number of perfect fluids.  Using finite volume and high-resolution shock-capturing methods, we dynamically evolve the two-fluid case, with the first fluid modeling ordinary matter and the second fluid modeling dark matter.  We use our dynamical solutions to study nonlinear stability, radial oscillation frequencies, and a dynamical formation process.
\end{abstract} 

\maketitle


\section{Introduction}

Dark matter could be mixed with ordinary matter inside neutron stars in such a way that dark matter affects bulk properties of the star, such as the mass and radius.  Measurements of neutron stars could then indirectly probe the properties of dark matter. Dark matter admixed neutron stars are mixtures of the ordinary nuclear matter of a neutron star and dark matter, with dark matter modeled as either a bosonic \cite{Henriques:1989ar, Henriques:1989ez, Henriques:1990xg, deSousa:1995ye, Pisano:1995yk, Sakamoto:1998aj, deSousa:2000eq, Henriques:2003yr, Dzhunushaliev:2011ma, ValdezAlvarado:2012xc, Brito:2015yga, Brito:2015yfh, Bezares:2019jcb, DiGiovanni:2020frc, Valdez-Alvarado:2020vqa, DiGiovanni:2021vlu, Kain:2021bwd, Karkevandi:2021ygv, Lee:2021yyn, DiGiovanni:2021ejn} or fermionic \cite{Sandin:2008db, Ciarcelluti:2010ji, Goldman:2011aa, Leung:2011zz, Leung:2012vea, Li_2012, Li:2012ii, Leung:2013pra, Goldman:2013qla, Xiang:2013xwa, Tolos:2015qra, Mukhopadhyay:2015xhs, Panotopoulos:2017pgv, Panotopoulos:2017idn, Gresham:2018rqo, Nelson:2018xtr, Ellis:2018bkr, Deliyergiyev:2019vti, Bhat:2019tnz, DelPopolo:2020pzh, Zhang:2020dfi, Das:2020vng, Kain:2020zjs, Kain:2021hpk, Das:2021dru, Das:2021yny, Sen:2021wev, Das:2021wku, Das:2020ptd, Jimenez:2021nmr} particle.  In the bosonic case, these systems are more generally known as fermion-boson stars, a name we shall use throughout, reserving dark matter admixed neutron stars for the fermionic case. 

Both systems have been studied extensively at the level of static solutions, which are solutions for which spacetime is time independent.  Fermion-boson stars have also been studied dynamically using full numerical relativity  \cite{ValdezAlvarado:2012xc, Brito:2015yga, Brito:2015yfh, Bezares:2019jcb, DiGiovanni:2020frc, Valdez-Alvarado:2020vqa, DiGiovanni:2021vlu}.  The main purpose of this article is to present for the first time a dynamical study of dark matter admixed neutron stars.  We do so by deriving the equations of motion, in conservation form, for spherically symmetric systems with an arbitrary number of perfect fluids.  We have developed a hydrodynamical code that uses finite volume and high-resolution shock-capturing methods to numerically solve the two-fluid system, where the first fluid models ordinary matter and the second fluid models fermionic dark matter.

How dark matter affects the mass and radius of a mixed star is well studied.  Less studied is how dark matter affects radial oscillation frequencies.  Although radial oscillations have yet to be measured for neutron stars \cite{Brillante:2014lwa, Sagun:2020qvc, Sun:2021cez}, they are a potentially valuable signature for the star.  The hope is that their study will reveal details about the stellar interior, such as the presence and properties of dark matter.  Radial oscillation frequencies may be computed by solving pulsation equations, which are derived from time dependent harmonic perturbations of static solutions, or by Fourier transforming dynamic solutions.  Pulsation equations have been derived for fermion stars \cite{Chandrasekhar:1964zz}, boson stars \cite{Gleiser:1988rq, Jetzer:1988vr, Gleiser:1988ih, Kain:2021rmk}, and for dark matter admixed neutron stars \cite{Comer:1999rs, Kain:2020zjs}, but not for fermion-boson stars.  On the other hand, as mentioned, fermion-boson stars have been dynamically evolved, from which radial oscillation frequencies have been computed using a Fourier transform \cite{ValdezAlvarado:2012xc, Valdez-Alvarado:2020vqa}.  Since pulsation equations and dynamical solutions have not previously been available for the same type of mixed star, a comparison of the two methods has never been made.  We compare the frequencies computed from the Fourier transform of our dynamical solutions with the frequencies computed from pulsation equations and find excellent agreement.

An import question is how dark matter could become mixed with ordinary matter in a neutron star.  One possibility is through capture, which can occur if there exists nongravitational interactions between ordinary matter and dark matter \cite{Goldman:1989nd, Kouvaris:2007ay, Bertone:2007ae, deLavallaz:2010wp, Kouvaris:2010vv, Brito:2015yga, Cermeno:2017xwb}.  Over the lifetime of a neutron star, however, the mass of the accumulated dark matter does not typically exceed $10^{-10}$ M$_\odot$ \cite{Goldman:1989nd, Kouvaris:2007ay, Ellis:2018bkr, Deliyergiyev:2019vti}, which has a negligible effect on the bulk properties of the star.  Another possibility is that mixing occurs during the stellar formation process.  Dynamical evolutions are likely the only theoretical method available for studying this possibility.  Since bosons exhibit gravitational cooling \cite{Seidel:1993zk, DiGiovanni:2018bvo}, in which a large percentage of bosons are able to dissipate to infinity, it does not appear too difficult for stable configurations of fermion-boson stars to grow out of stable fermion stars enveloped in clouds of bosonic dark matter \cite{DiGiovanni:2020frc}.  Less clear are the prospects for the formation of dark matter admixed neutron stars, since fermions do not exhibit gravitational cooling.  We show that a stable fermion star enveloped by fermionic dark matter can lead to a stable dark matter admixed neutron star.

In the next section, we present our model.  In particular, we present the conservative variables we will be using, derive the equations of motion in conservation form, and discuss our numerical methods.  We also review solving for static solutions and discuss the equations of state we use for both ordinary and dark matter.  In Sec.\ \ref{sec:1 fluid}, we present results for neutron stars (i.e.\ single-fluid stars).  We use this single-fluid case as a warm-up, focusing on nonlinear stability and radial oscillation frequencies in a more familiar setting.  In Sec.\ \ref{sec:dmans}, we study dark matter admixed neutron stars (i.e.\ two-fluid stars) and present our primary results.  We compute the (linear) stability curve over the whole of parameter space and present results for nonlinear stability.  We then study radial oscillation frequencies, showing that our dynamical solutions and our solutions to pulsation equations agree.  Finally, we present results for a dynamical formation process.  We conclude in Sec.\ \ref{sec:conclusion}.


\section{Multiple-fluid stars}

In this section, we derive equations that describe spherically symmetric stars made up of an arbitrary number of perfect fluids.  We present these equations in conservation form, after introducing conservative variables, which allows us to solve them using finite volume and high-resolution shock-capturing methods.  

We parametrize the spherically symmetric metric as
\begin{equation} \label{metric}
ds^2 = -\alpha^2 dt^2 + a^2 dr^2 + r^2(d\theta^2 + \sin^2\theta d\phi^2),
\end{equation}
with metric functions $\alpha(t,r)$ and $a(t,r)$, and use units such that $c = \hbar = 1$.  We make the standard assumption that nongravitational interactions between ordinary and dark matter are negligible.  As a consequence, the energy-momentum tensor separates, so that it may be written as
\begin{equation} \label{T tot}
(T_\text{tot})^{\mu\nu} = \sum_x (T_x)^{\mu\nu},
\end{equation}
where $x$ labels the fluid,
\begin{equation} \label{Tx}
(T_x)^{\mu\nu} = (\rho_x + P_x)u_x^\mu u_x^\nu + P_x g^{\mu\nu}
\end{equation}
is the energy-momentum tensor for a perfect fluid, $u_x^\mu$ is the four-velocity of the fluid, and $\rho_x$ and $P_x$ are the fluid's energy density and pressure.  $(T_\text{tot})^{\mu\nu}$ in Eq.\ (\ref{T tot}) is the (total) energy-momentum tensor of the system and is what goes on the right-hand side of the Einstein field equations,
\begin{equation} \label{GR eq}
G^{\mu\nu} = 8\pi G (T_\text{tot})^{\mu\nu},
\end{equation}
where $G^{\mu\nu}$ is the Einstein tensor and $G$ is the gravitational constant.  In addition to the energy-momentum tensor, the matter sector is defined by equations of state, which we also assume separate,
\begin{equation} \label{eos}
P_x = P_x (\rho_x),
\end{equation}
i.e.\ $P_x$ depends only on its associated energy density $\rho_x$ and not on any $\rho_{y\neq x}$.  We stress that we are not allowing for the more general form $P_x = P_x(\rho_x, n_x)$, where $n_x$ is the fluid number density; equations of state of the form (\ref{eos}) are called barotropic.  Our assumptions that the energy-momentum tensor and the equations of state separate means that there are only gravitational interfluid interactions and that the individual $(T_x)^{\mu\nu}$ are conserved, $\nabla_\mu (T_x)^{\mu\nu} = 0$, in addition to the requisite $\nabla_\mu (T_\text{tot})^{\mu\nu} = 0$.

Equations that determine the metric functions $\alpha$ and $a$ follow from the Einstein field equations and are
\begin{equation} \label{metric equations}
\begin{split}
\frac{\partial_r \alpha}{\alpha} &= +4\pi G r  a^2 (T_\text{tot})\indices{^r_r} + \frac{a^2-1}{2r}
\\
\frac{\partial_r a}{a} &= 
-4\pi G r a^2 (T_\text{tot})\indices{^t_t}
- \frac{a^2 - 1}{2r} 
\\
\frac{\partial_t a}{a} &= -4\pi G r \alpha^2 (T_\text{tot})\indices{^t_r}.
\end{split}
\end{equation}


\subsection{Primitive and conservative variables}

In spherical symmetry the fluid velocity must be zero in the $\theta$- and $\phi$-directions and thus $u_x^\theta = u_x^\phi = 0$.  Defining
\begin{equation} \label{W v def}
W_x \equiv \alpha u_x^t, \qquad
v_x \equiv \frac{a u_x^r}{\alpha u_x^t} = \frac{a u_x^r}{W_x},
\end{equation}
it follows from $u_x^\mu u^x_\mu = -1$ that $W_x^2 = 1/(1-v_x^2)$ is the relativistic factor.  $P_x$, $\rho_x$, $v_x$, and $W_x$ are collectively known as the \textit{primitive} variables.  After specifying an equation of state, $P_x(\rho_x)$, and noting that $W_x$ is a function of $v_x$, the matter sector is completely described by $\rho_x$ and $v_x$.

To solve for the primitive variables we will write the equations of motion in \textit{conservation form}, which requires the introduction of \textit{conservative} variables.  To introduce the standard set of conservative variables, we first introduce the rest mass energy density, 
\begin{equation} \label{rhox0}
\rho_{x}^\text{rest} = m_x n_x,
\end{equation}
where $m_x$ is the fluid rest mass and, as mentioned above, $n_x$ is the fluid number density.  The standard set of conservative variables is \cite{Romero:1995cn, RezzollaBook}
\begin{equation} \label{D S E}
\begin{split}
D_x &\equiv a \rho_{x}^\text{rest} W_x
\\
S_x &\equiv (\rho_x + P_x)W_x^2 v_x
\\
E_x &\equiv  (\rho_x + P_x)W_x^2 - P_x,
\end{split}
\end{equation}
after which one replaces replaces $E_x$ with $\tau_x \equiv E_x - D_x$, which is more accurate numerically since it is the combination of two conservative variables \cite{RezzollaBook}.  Although this is the standard set, it is not the set we will be using, as will be made clear in Sec.\ \ref{sec:new vars}.  In terms of the standard set, the nonvanishing components of the energy-momentum tensor are 
\begin{equation} \label{T D S Tau}
\begin{split}
(T_x)\indices{^t_t} &= -E_x
\\
(T_x)\indices{^t_r} &=  \frac{a}{\alpha} S_x
\\
(T_x)\indices{^r_t} &=  - \frac{\alpha}{a} S_x
\\
(T_x)\indices{^r_r} &= S_x v_x + P_x
\\
(T_x)\indices{^\theta_\theta} &=
(T_x)\indices{^\phi_\phi} = P_x.
\end{split}
\end{equation}


\subsection{Equations of motion}

Equations of motion follow from conservation of the individual energy-momentum tensors, $\nabla_\mu (T_x)\indices{^\mu_\nu} = 0$.  Using the metric in (\ref{metric}), the $\nu = t,r$ equations work out to
\begin{align} 
0 &= \partial_t (T_x)\indices{^t_t} + \partial_r (T_x)\indices{^r_t}
\notag\\
&\qquad
+ \frac{\dot{a}}{a} [(T_x)\indices{^t_t} - (T_x)\indices{^r_r}]
+ \left( \frac{\alpha'}{\alpha} + \frac{a'}{a} + \frac{2}{r} \right) (T_x)\indices{^r_t}
\notag\\
0 &= \partial_t (T_x)\indices{^t_r} + \partial_r (T_x)\indices{^r_r}
+ \left( \frac{\dot{\alpha}}{\alpha} + \frac{\dot{a}}{a} \right) (T_x)\indices{^t_r}
\notag\\
&\qquad
+ \frac{\alpha'}{\alpha} [(T_x)\indices{^r_r} - (T_x)\indices{^t_t}]
\notag\\
&\qquad
+ \frac{1}{r} [2 (T_x)\indices{^r_r} - (T_x)\indices{^\theta_\theta} - (T_x)\indices{^\phi_\phi}],
\label{eom}
\end{align}
where a dot denotes a $t$-derivative and a prime denotes an $r$-derivative.  We also have an equation of motion from conservation of particle number, $\nabla_\mu (n_x u_x^\mu)$ = 0, which is often written as $\nabla_\mu (\rho_{x}^\text{rest} u_x^\mu) = 0$ using (\ref{rhox0}).  Using the metric in (\ref{metric}) and the definition of $W_x$ and $v_x$ in (\ref{W v def}), this becomes
\begin{equation} \label{num eom}
\begin{split}
0 &=
\partial_t \left(\frac{\rho_{x}^\text{rest} W_x}{\alpha} \right) 
+ \frac{\rho_{x}^\text{rest} W_x v_x}{a} \left( \frac{\alpha'}{\alpha} + \frac{a'}{a} + \frac{2}{r} \right)
\\
&\qquad
+ \partial_r \left(\frac{\rho_{x}^\text{rest} W_x v_x}{a} \right)
+  \frac{\rho_{x}^\text{rest} W_x }{\alpha} \left(\frac{\dot{\alpha}}{\alpha} + \frac{\dot{a}}{a} \right)
.
\end{split}
\end{equation}

To facilitate solving these equations of motion, we write them in conservation form.  We are not aware of this having been done before for an arbitrary number of perfect fluids.  In conservation form, the equations of motion are
\begin{equation} \label{consv eom}
\partial_t \mathbf{u}_x + \frac{1}{r^2} \partial_r \left(r^2 \frac{\alpha}{a} \mathbf{f}_x \right) = \mathbf{s}_x,
\end{equation} 
where $\textbf{u}_x$ is the vector of conservative variables, $\textbf{f}_x$ is the vector of fluxes, and $\textbf{s}_x$ is the vector of sources,
\begin{equation} \label{conservation form}
\mathbf{u}_x = 
\begin{pmatrix}
D_x \\ S_x \\ E_x
\end{pmatrix}
\quad
\mathbf{f}_x = 
\begin{pmatrix}
D_x v_x \\ S_x v_x + P_x \\ S_x
\end{pmatrix}
\quad
\mathbf{s}_x = 
\begin{pmatrix}
s_x^{D} \\ s_x^{S} \\ s_x^{E}
\end{pmatrix},
\end{equation}
with the sources given by
\begin{align} 
s_x^{D} &= 0
\notag\\
s_x^S &= 4\pi G r a \alpha \sum_y ( 2 S_y v_x P_x - E_x P_y - E_y P_x )
\notag\\
&\qquad
+ \frac{\alpha(a^2-1) }{2r a} (S_x v_x - E_x + P_x) 
+ \frac{2\alpha}{ra} P_x
\notag\\
&\qquad 
+ 4\pi G r a \alpha \sum_y 
(2 E_x v_x S_y - E_x S_y v_y - E_y S_x v_x )
\notag\\
s_x^E &= 
4\pi G r a \alpha \sum_y
\Bigl[S_y (E_x + S_x v_x + P_x) 
\notag\\
&\qquad\qquad\qquad\quad
- S_x (E_y + S_y v_y + P_y) \Bigr].
\label{sources}
\end{align}
For a single fluid, these reduce down to the well-known formulas \cite{Romero:1995cn, RezzollaBook}.  It is straightforward to replace the $E_x$ equation of motion with that for $\tau_x = E_x - D_x$.


\subsection{New conservative variables}
\label{sec:new vars}

The equations of motion for $S_x$ and $E_x$ in (\ref{consv eom}) are independent of $D_x$.  Further, $D_x$ is the only conservative variable that depends on the rest mass energy density $\rho_{x}^\text{rest}$, as can be seen from (\ref{D S E}).  We mentioned above that we are assuming barotropic equations of state, $P_x(\rho_x)$, and hence our equations of state do not explicitly depend on $\rho_{x}^\text{rest}$.  As a consequence, $D_x$ decouples and we do not have to solve for it.

Following Neilsen and Choptuik \cite{Neilsen:1999we}, we introduce new conservative variables
\begin{equation} \label{Pi Phi}
\Pi_x \equiv E_x+S_x,
\qquad
\Phi_x \equiv E_x-S_x,
\end{equation}
which we use instead of $S_x$ and $E_x$, and drop $D_x$.  In terms of the primitive variables, the new conservatives variables are given by
\begin{equation} \label{con from prim}
\Pi_x = \frac{\rho_x + P_x}{1-v_x} - P_x, \qquad
\Phi_x = \frac{\rho_x + P_x}{1+v_x} - P_x,
\end{equation}
which follows from Eqs.\ (\ref{D S E}) and (\ref{Pi Phi}).

In presenting the equations of motion in terms of the new conservative variables, we separate the flux into two terms,
\begin{equation}
\mathbf{f}_x = \mathbf{f}_x^{(1)} + \mathbf{f}_x^{(2)},
\end{equation}
so that the equations of motion become
\begin{equation} \label{eom 2}
\partial_t \mathbf{u}_x + \frac{1}{r^2} \partial_r \left(r^2 \frac{\alpha}{a} \mathbf{f}_x^{(1)} \right)
+ \partial_r \left(\frac{\alpha}{a} \mathbf{f}_x^{(2)} \right)  = \mathbf{s}_x,
\end{equation} 
where now
\begin{align} \label{conservation form 2}
\mathbf{u}_x &= 
\begin{pmatrix}
\Pi_x \\ \Phi_x
\end{pmatrix}& 
\mathbf{f}_x^{(1)} &= 
\begin{pmatrix}
\frac{1}{2}(\Pi_x - \Phi_x)( 1 + v_x) \\
\frac{1}{2}(\Pi_x - \Phi_x)( 1 - v_x)
\end{pmatrix}
\notag \\
\mathbf{f}_x^{(2)} &= 
\begin{pmatrix}
+ P_x
\\
- P_x
\end{pmatrix}&
\mathbf{s}_x &= 
\begin{pmatrix}
\Omega_x + \Theta_x
\\
\Omega_x - \Theta_x
\end{pmatrix}
\end{align}
and
\begin{equation} \label{Omega Theta}
\begin{split}
\Omega_x &\equiv
4 \pi G r a \alpha \sum_y
\Bigl[S_y (E_x + S_x v_x + P_x) 
\\
&\qquad\qquad\qquad\quad
- S_x (E_y + S_y v_y + P_y) \Bigr]
\\
\Theta_x &\equiv
4\pi G r a \alpha \sum_y ( 2 S_y v_x P_x - E_x P_y - E_y P_x )
\\
&\quad
+ \frac{\alpha(a^2-1)}{2r a} (S_x v_x - E_x + P_x) 
\\
&\quad
+ 4\pi G r a \alpha \sum_y 
(2 E_x v_x S_y - E_x S_y v_y - E_y S_x v_x ).
\end{split}
\end{equation}
In $\Omega_x$ and $\Theta_x$, $E_x$ and $S_x$ are to be replaced with
\begin{equation}
E_x = \frac{1}{2} \left( \Pi_x + \Phi_x \right), 
\qquad
S_x = \frac{1}{2} \left( \Pi_x - \Phi_x \right).
\end{equation}
The reason for separating the flux into two terms is that it leads to the cancellation of the $2\alpha P_x/ra$ term in $s_x^S$ in Eq.\ (\ref{sources}).  This cancellation can be difficult to achieve precisely numerically, and so the numerical solution is improved if the cancellation is performed at the level of the equations.

At each integration step, the solution to the equations of motion gives the conservative variables $\Pi_x$ and $\Phi_x$.  To perform the next integration step, we must solve for the primitive variables, since both the fluxes and sources contain them, as can be seen in Eqs.\ (\ref{conservation form 2}) and (\ref{Omega Theta}).  Unfortunately, it is not possible to find a general analytical formula for doing so and it must be done numerically.  Trying to invert (\ref{con from prim}) leads to
\begin{align} \label{prim from cons}
0 &= (\Pi_x - \Phi_x)^2 - (\Pi_x + \Phi_x - 2\rho_x)(\Pi_x + \Phi_x + 2P_x)
\notag \\
v_x &= \frac{\Pi_x - \Phi_x}{\Pi_x + \Phi_x + 2P_x}.
\end{align}
When combined with the equation of state $P_x(\rho_x)$, the first equation can be solved numerically for $\rho_x$, which gives $P_x$, from which the second equation gives $v_x$.

We end this subsection listing the metric variable equations in (\ref{metric equations}), but written in terms of the new conservative variables,
\begin{align} \label{metric eqs}
\frac{\partial_r \alpha}{\alpha} &= 4\pi G r  a^2 \sum_y \left( \frac{\Pi_y - \Phi_y}{2}  v_y + P_y \right) + \frac{a^2-1}{2r}
\notag \\
\frac{\partial_r a}{a} &= 
4\pi G r a^2 \sum_y \frac{\Pi_y + \Phi_y}{2}
- \frac{a^2 - 1}{2r} 
\notag \\
\frac{\partial_t a}{a} &= -4\pi G r \alpha a \sum_y \frac{\Pi_y - \Phi_y}{2}.
\end{align}


\subsection{Numerical methods}
\label{sec:numerical}

Dynamical evolution of hydrodynamical systems is hampered by the generic formation of discontinuities, or shocks.  This occurs even for smooth initial data and causes the failure of simple finite difference schemes.  We overcome this by using finite volume and high-resolution shock-capturing methods (see, for example, \cite{Romero:1995cn, RezzollaBook, Neilsen:1999we, Guzman:2012rn}).  Finite volume methods write the equations of motion in Eq.\ (\ref{eom 2}) in an integral form, which can handle shocks.  In effect, we introduce a uniform computational grid, where each grid point sits at the center of a cell, and replace the variables in  Eq.\ (\ref{eom 2}) with their cell-averaged values.  We then solve the equations of motion using the method of lines, with second order finite differencing used for spatial derivatives,
\begin{align} 
\partial_t\bar{\mathbf{u}}_{x,i} &= 
- \frac{3 \left[ \left(r^2 \frac{\alpha}{a} \mathbf{f}_x^{(1)} \right)_{i+1/2} - \left(r^2 \frac{\alpha}{a} \mathbf{f}_x^{(1)} \right)_{i-1/2} \right]}
{r^3_{i+1/2} - r^3_{i-1/2}}
\notag \\
&\qquad
- \frac{1}{\Delta r}  \left[ \left(\frac{\alpha}{a} \mathbf{f}_x^{(2)} \right)_{i+1/2} - \left(\frac{\alpha}{a} \mathbf{f}_x^{(2)} \right)_{i-1/2} \right]
\notag \\
&\qquad
+ \bar{\mathbf{s}}_{x,i}.
\end{align}
The $\bar{\mathbf{u}}_{x,i}$ are the cell averaged conservative variables and $\bar{\mathbf{s}}_{x,i} = \mathbf{s}_x(\bar{\mathbf{u}}_i)$.  The subscript $i$ indicates the center of a cell (i.e.\ a grid point), while $i\pm 1/2$ indicate cell boundaries.  The finite differencing of the first spatial derivative in Eq.\ (\ref{eom 2}) makes use of the standard technique $r^{-2} \partial_r \rightarrow 3\partial_{r^3}$.

An immediate difficulty arises in how to evaluate the fluxes at cell boundaries.  The $\bar{\mathbf{u}}_{x,i}$, which are used to evaluate the fluxes, are cell averaged values and thus are discontinuous across cell boundaries.  This is known as a local Riemann problem and we handle it using Godunov's method.  We use the minmod slope limiter to perform a linear reconstruction at each cell boundary and then use an approximate Riemann solver to solve the local Rimeann problem for the fluxes.  We have implemented both the HLLE and Roe solvers, finding negligible differences between results.  These solvers make use of the spectral decomposition of the Jacobian matrix $A \equiv \partial \mathbf{f}/\partial \mathbf{u}$.  Note that the flux, $\mathbf{f}_x$, does not mix fluids and only depends on the properties of fluid $x$, as can be seen in Eq.\ (\ref{conservation form 2}).  As a consequence, we can write the Jacobian matrix as
\begin{equation} \label{Jacobian}
A_x = \frac{\partial \mathbf{f}_x}{\partial \mathbf{u}_x}.
\end{equation}
Its spectral decomposition is given in Appendix \ref{app:Jacobian}.

We use third-order Runge-Kutta \cite{SHU1988439} to solve the equations of motion in the time direction.  At each integration step, we numerically solve the top equation in (\ref{prim from cons}) for $\rho_x$ using the Newton-Raphson method.  For metric variables, we solve the top two equations in (\ref{metric eqs}) using second-order Runge-Kutta, the inner boundary condition $a = 1$, and the outer boundary condition $\alpha = 1/a$.   The remaining numerical methods we use can be found in \cite{Neilsen:1999we}, including additional boundary conditions and the implementation of a floor.

Our code uses dimensionless time and space variables defined by
\begin{equation}
\bar{t} \equiv \sqrt{G} (1\text{ GeV})^2 \, t, \qquad
\bar{r} \equiv \sqrt{G} (1\text{ GeV})^2 \, r.
\end{equation}
The results shown in upcoming sections were computed using uniform grid spacing $\Delta \bar{r} = 0.01$, time step $\Delta \bar{t} = 0.5 \Delta \bar{r}$,  and an outer boundary at $\bar{r}_\text{max} = 350$.  We show in Appendix \ref{app:code tests} that our code is second order convergent.


\subsection{Initial data:\ Static solutions}

We are interested in making a dynamical study of dark matter admixed neutron stars.  This system has typically been studied through static solutions, which are solutions for which all fields are time independent.  By using static solutions as initial data, our dynamical evolutions can determine their nonlinear stability.  For unstable solutions, we can further determine what the static solutions evolve to and for stable solutions, we can compute the radial oscillation frequencies by Fourier transforming the dynamic solution.

Time independent fields require vanishing fluid velocities, which reduces the individual energy-momentum tensors to
\begin{equation} \label{static T}
(T_x)\indices{^\mu_\nu} = \text{diag}(-\rho_x, P_x, P_x, P_x).
\end{equation}
It is convenient to replace the metric field $a$ with
\begin{equation} \label{m def}
m = \frac{r}{2G} \left(1- \frac{1}{a^2} \right),
\end{equation}
which gives the total mass inside a radius $r$.  Using Eqs.\ (\ref{static T}) and (\ref{m def}), the first and second metric equations in (\ref{metric equations}), and the equations of motion $\nabla_\mu (T_x)\indices{^\mu_\nu} = 0$, we have
\begin{equation} \label{TOV}
\begin{split}
\partial_r \alpha &= \alpha G \frac{4\pi r^3 \sum_y P_y + m}{r^2(1-2Gm/r)}
\\
\partial_r m &= 4\pi r^2 \sum_y \rho_y
\\
\partial_r P_x &= - G \frac{4\pi r^3 \sum_y P_y + m}{r^2(1-2Gm/r)} (\rho_x + P_x),
\end{split}
\end{equation}
which are the multifluid Tolman--Oppenheimer--Volkoff (TOV) equations.  The solution to these equations are the static solutions we use as initial data.

Static solutions are uniquely identified by the central pressures $P_x(0)$.  To solve the TOV equations, we specify the $P_x(0)$ and integrate the $m$ and $P_x$ equations in (\ref{TOV}) outward from $r = 0$ using the inner boundary condition $m(0) = 0$.  The edge of fluid $x$, at $r = R_x$, is defined by $P_x(R_x) = 0$ and the edge of the star, at $r = R$, is given by the outermost $R_x$.  The mass of the star is given by $M = m(R)$.  Outside the star, the spacetime is Schwarzschild and thus $\alpha = 1/a = \sqrt{1 - 2Gm/r}$ for $r \geq R$.  Using this as the outer boundary condition, the $\alpha$ equation in (\ref{TOV}) can be integrated inward from $r=R$.


\subsection{Equations of state}
\label{sec:eos}

Our model allows for arbitrary equations of state of the form $P_x(\rho_x)$, as long as they are sufficiently smooth so as to maintain numerical stability.  Ideally, for ordinary matter, we would use a realistic equation of state, one that takes into account charge neutrality, beta-equilibrium, and inner and outer crusts.  Unfortunately, such equations of state are typically presented in a tabulated form, rendering them insufficiently smooth.  All dynamical evolutions of fermion-boson stars that we are aware of \cite{ValdezAlvarado:2012xc, Brito:2015yga, Brito:2015yfh, Bezares:2019jcb, DiGiovanni:2020frc, Valdez-Alvarado:2020vqa, DiGiovanni:2021vlu} have used a simple polytropic equation of state for ordinary matter,
\begin{equation} \label{poly eos}
P_x = K \rho_x^\gamma,
\end{equation}
with $\gamma = 2$ and $K = 100$ GeV$^{-4}$, and we do the same here.  Although this is not a realistic equation of state, we expect it to capture qualitative aspects of the ordinary matter in a neutron star.   We are currently investigating the possibility of evolving realistic equations of state and will report results elsewhere.

We use the same equation of state for both ordinary and dark matter.  We do this in part for simplicity, but there is also a physical motivation.  Mirror dark matter follows from the assumption that the Universe is parity symmetric.  The addition of new particles to restore parity to the Standard Model leads to mirror baryons as viable candidates for dark matter (see \cite{Okun:2006eb, Hippert:2021fch} and references therein).  For dark matter admixed neutron stars with mirror dark matter, it is customary to use the \textit{same} equation of state for ordinary and dark matter \cite{Sandin:2008db, Ciarcelluti:2010ji, Goldman:2011aa, Goldman:2013qla}.


\section{Single-fluid stars}
\label{sec:1 fluid}

In this section, we present results for single-fluid stars, i.e.\ for neutron stars.  These results will help us better understand the analogous two-fluid results for dark matter admixed neutron stars given in the next section.  They also act as a nontrivial check, in a more familiar setting, on our numerical methods.

\begin{figure}
\centering
\includegraphics[width=2.5in]{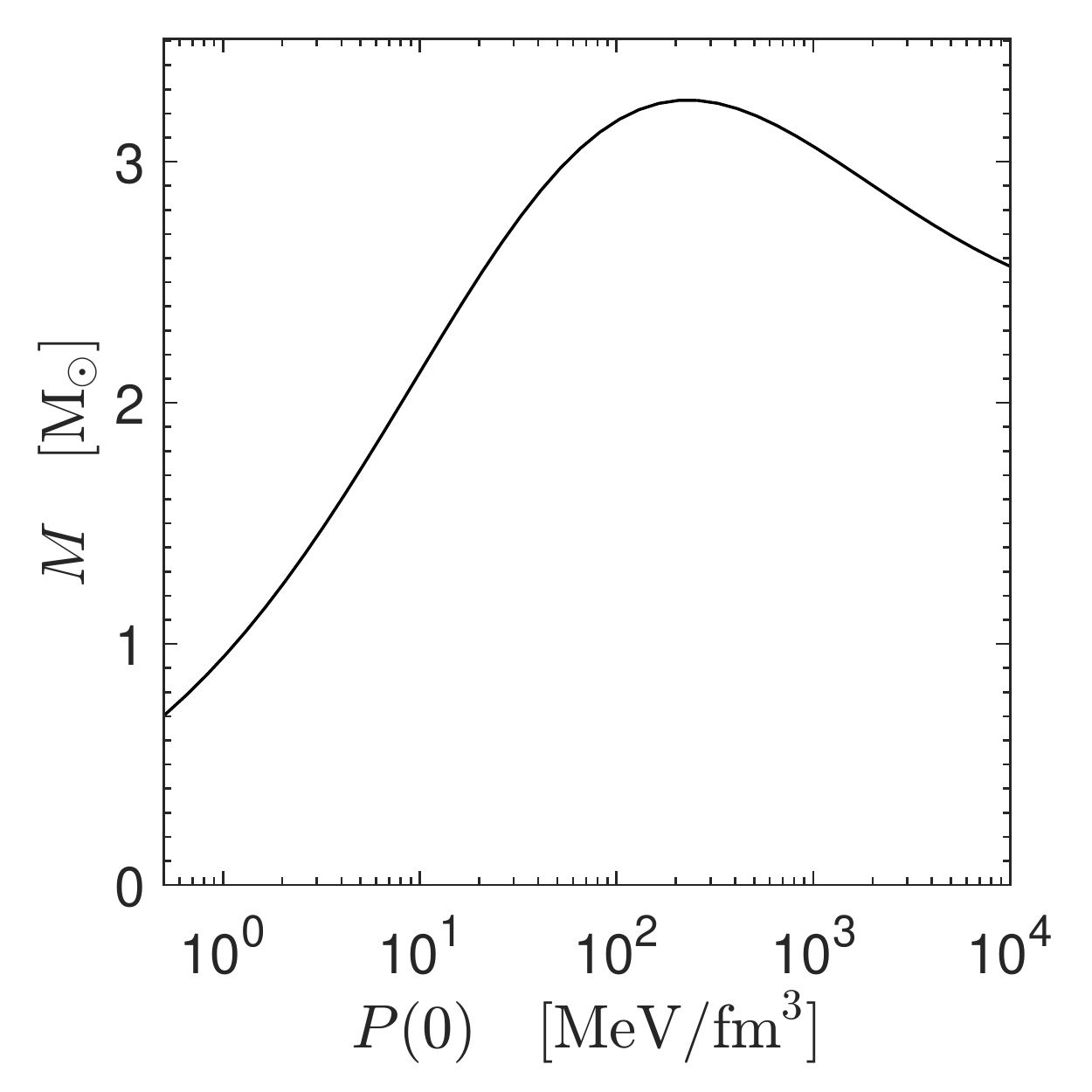}
\caption{The mass as a function of the central pressure for single-fluid stars with the equation state in Eq.\ (\ref{poly eos}).  Each point on the curve represents a static solution.}
\label{fig:1 fluid}
\end{figure}

In Fig.\ \ref{fig:1 fluid}, we show solutions to the single-fluid TOV equations for a range of central pressures for the polytropic equation of state.  Each point on the curve represents a static solution.  The static solution with the largest mass is called the critical solution, which has the critical central pressure
\begin{equation} \label{1 fluid crit}
P(0) = 230.4 \text{ MeV/fm}^3
\end{equation}
and corresponding critical mass $M = 3.255$ M$_\odot$.

Static solutions with central pressures smaller than the critical pressure are linearly stable, otherwise they are unstable.  One way to derive this result is to perturb a static solution with time dependent harmonic perturbations, which
depend on the radial oscillation frequency.  Upon linearizing the equations, one may derive a pulsation equation, whose solution gives the squared radial oscillation frequency.  The pulsation equation has a discretely infinite number of solutions, each with a distinct frequency.  The solution with the smallest frequency is called the fundamental solution and if the squared radial oscillation frequency is positive for the fundamental solution, the corresponding static solution is linearly stable, otherwise it is unstable.  The single-fluid pulsation equation was derived some time ago by Chandrasekhar \cite{Chandrasekhar:1964zz}.  Its multifluid generalization \cite{Kain:2020zjs} is reviewed in Appendix \ref{app:pulsation}.

Radial oscillation frequencies have been computed for various equations of state (see, for example, \cite{Glass1983, Gondek:1997fd, Kokkotas:2000up}) and they present an interesting signature for a neutron star.  The hope is that their study can reveal details about the inner structure of the star, including the equation of state.  Being spherically symmetric, radial oscillation modes do not couple to gravitational waves and it is unlikely they can be measured by purely gravitational means.  The expectation is that they could be measured by emission of electromagnetic radiation from charge that has accumulated on the surface of the star (see, for example, \cite{Brillante:2014lwa, Sagun:2020qvc, Sun:2021cez}).

\begin{figure}
\centering
\includegraphics[width=3.2in]{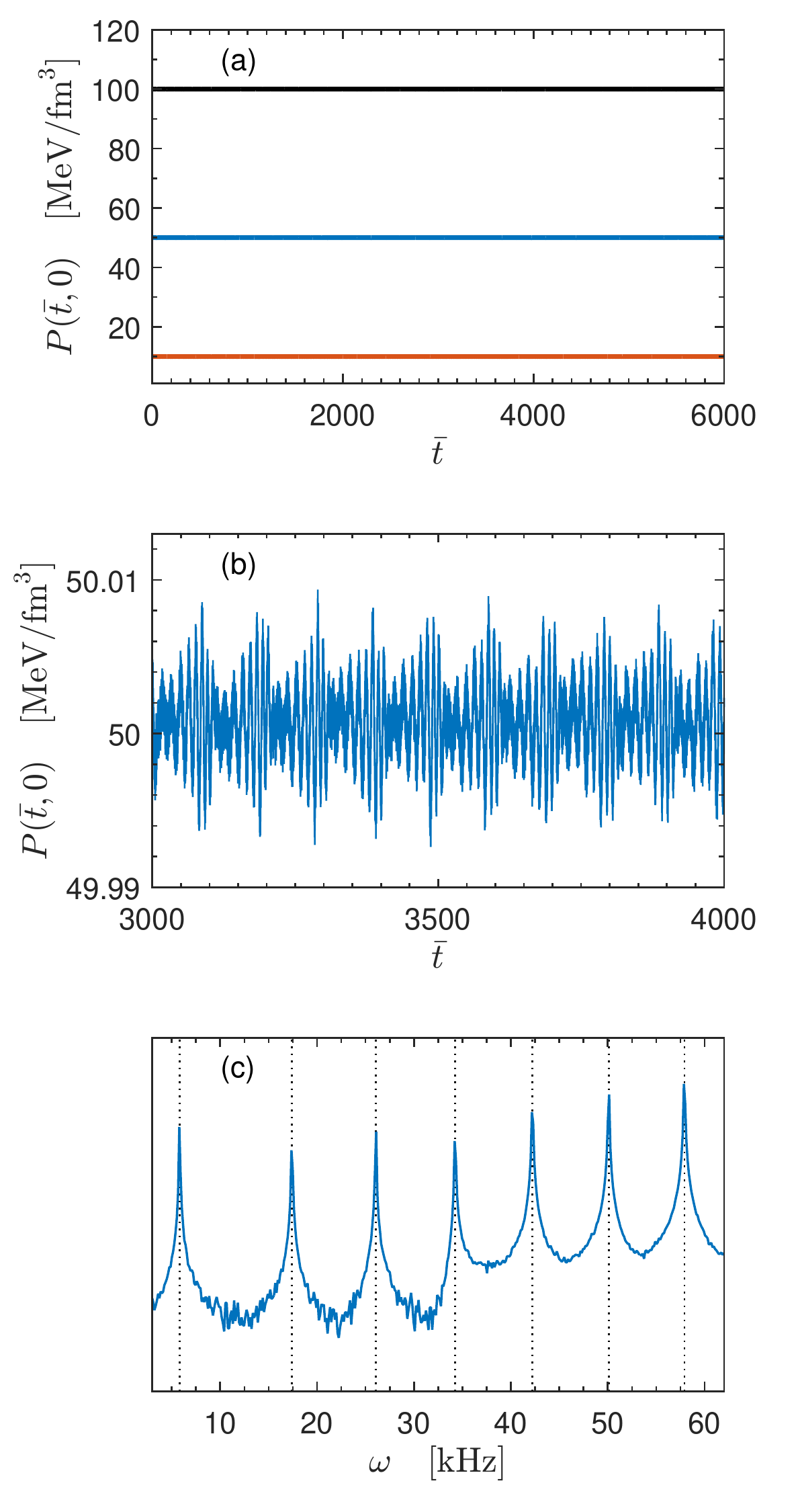}
\caption{(a) The central pressure as a function of time for three evolutions.  From top to bottom, the static solutions used as initial data are defined by $P(0) = 100$, $50$, and $10$ MeV/fm$^3$.  That the three lines remain straight indicates that these three linearly stable static solutions are also nonlinearly stable.  (b) Zooming in on the middle curve in (a) shows radial oscillations.  (c) The blue curve gives the Fourier transform of the middle curve in (a), with the spikes indicating the radial oscillation frequencies (the units for the vertical axis are arbitrary).  The dotted vertical lines are independent computations of the radial oscillation frequencies as computed from a pulsation equation.}
\label{fig:1 fluid FT}
\end{figure}

We have dynamically evolved single-fluid stars using static solutions as initial data.  As is well known, discretization error acts as a perturbation, inducing radial oscillations.  In Fig.\ \ref{fig:1 fluid FT}(a), we show the central pressure, $P(\bar{t},0)$, for three evolutions out to $\bar{t} = 6000$.  The static solutions used for these evolutions are defined by $P(0,0) = 100$ (top, black), 50 (middle, blue), and 10 MeV/fm$^3$ (bottom, orange), which are linearly stable as can be seen from Fig.\ \ref{fig:1 fluid}.  That the curves remain straight over the course of the evolutions indicates that these static solutions are also nonlinearly stable.  We find similar results for other linearly stable stars we evolved.  For all unstable stars we  evolved, the evolution collapsed to a black hole, which was indicated by a spike in the metric function $a$ and the collapse of $\alpha$ around $r=0$.  

Each of the curves in Fig.\ \ref{fig:1 fluid FT}(a) contain radial oscillations.  These oscillations become apparent upon zooming in, which we show in Fig.\ \ref{fig:1 fluid FT}(b) for the middle curve in Fig.\  \ref{fig:1 fluid FT}(a).  The fast Fourier transform of the middle curve in Fig.\ \ref{fig:1 fluid FT}(a) is shown in Fig.\ \ref{fig:1 fluid FT}(c).  The blue curve is the Fourier transform, with the spikes giving the radial oscillation frequencies.  The dotted vertical lines are independent computations of the radial oscillation frequencies as computed from the pulsation equation.  We can see excellent agreement between the two methods, which is a nontrivial check on our equations and code.  (For a different analysis see \cite{Gabler:2009yt}.)


\section{Dark matter admixed neutron stars}
\label{sec:dmans}

Dark matter admixed neutron stars are two-fluid systems, with one fluid describing the ordinary matter in a neutron star and the second fluid describing dark matter.  As explained in Sec.\ \ref{sec:eos}, we are using the same polytropic equation of state in Eq.\ (\ref{poly eos}) for both ordinary and dark matter.  In this section, we study dark matter admixed neutron stars using our two-fluid hydrodynamical code.


\subsection{Stability}

For single-fluid systems, static solutions are identified by the central pressure of the fluid and the transition from linearly stable to unstable is marked by a point in parameter space, which is called the critical point, as explained in Sec.\ \ref{sec:1 fluid}.  In two-fluid systems, static solutions are identified by the central pressures of each fluid and the transition from linearly stable to unstable is marked by a curve in parameter space, which is called the critical curve.

There are multiple ways to compute the critical curve.  One method is to perturb static solutions with time dependent harmonic perturbations, which depend on the radial oscillation frequency.  Upon linearizing the equations, one may derive a system of pulsation equations whose solution gives the squared radial oscillation frequency.  Just as in the single-fluid case, the two-fluid pulsation equations are expected to have an infinite number of solutions, each with a distinct frequency, and if the fundamental squared frequency is positive, the static solution is linearly stable, otherwise it is unstable.  Multifluid pulsation equations are reviewed in Appendix \ref{app:pulsation}.

An alternative approach was first presented in \cite{Henriques:1990xg}.  The critical curve is given by the solution to
\begin{equation} \label{stability conditions}
\frac{d M}{d\mathbf{p}} = 
\frac{d N_\text{om}}{d\mathbf{p}} = 
\frac{d N_\text{dm}}{d\mathbf{p}} = 0,
\end{equation}
where $M$ and $N_x$ are the total mass and fluid number of a static solution and $\mathbf{p}$ is a vector in parameter space that is simultaneously tangent to the level curves of $M$ and $N_x$.  It can be shown that if two of the quantities in (\ref{stability conditions}) are zero, then the third is also  \cite{Henriques:1990xg, Jetzer:1990xa}. In practice, we compute contour lines of either $N_\text{om}$ or $N_\text{dm}$.  Moving along a single contour line, we determine the point where $M$ is an extremum (in practice, we find that it is a maximum).  These points give the critical curve.  (For a derivation of this result, as well as the equations for computing $N_\text{om}$ and $N_\text{dm}$ from an equation of state, see \cite{Kain:2021hpk}.)

The first method in terms of harmonic perturbations has the advantage that it can compute radial oscillation frequencies, but it is a time consuming method for computing a critical curve.  The second method using Eq.\ (\ref{stability conditions}) is much faster, but it cannot compute radial oscillation frequencies.  It was confirmed in \cite{Kain:2021hpk} that these two methods give the same result for critical curves, as expected.

\begin{figure}
\centering
\includegraphics[width=3in]{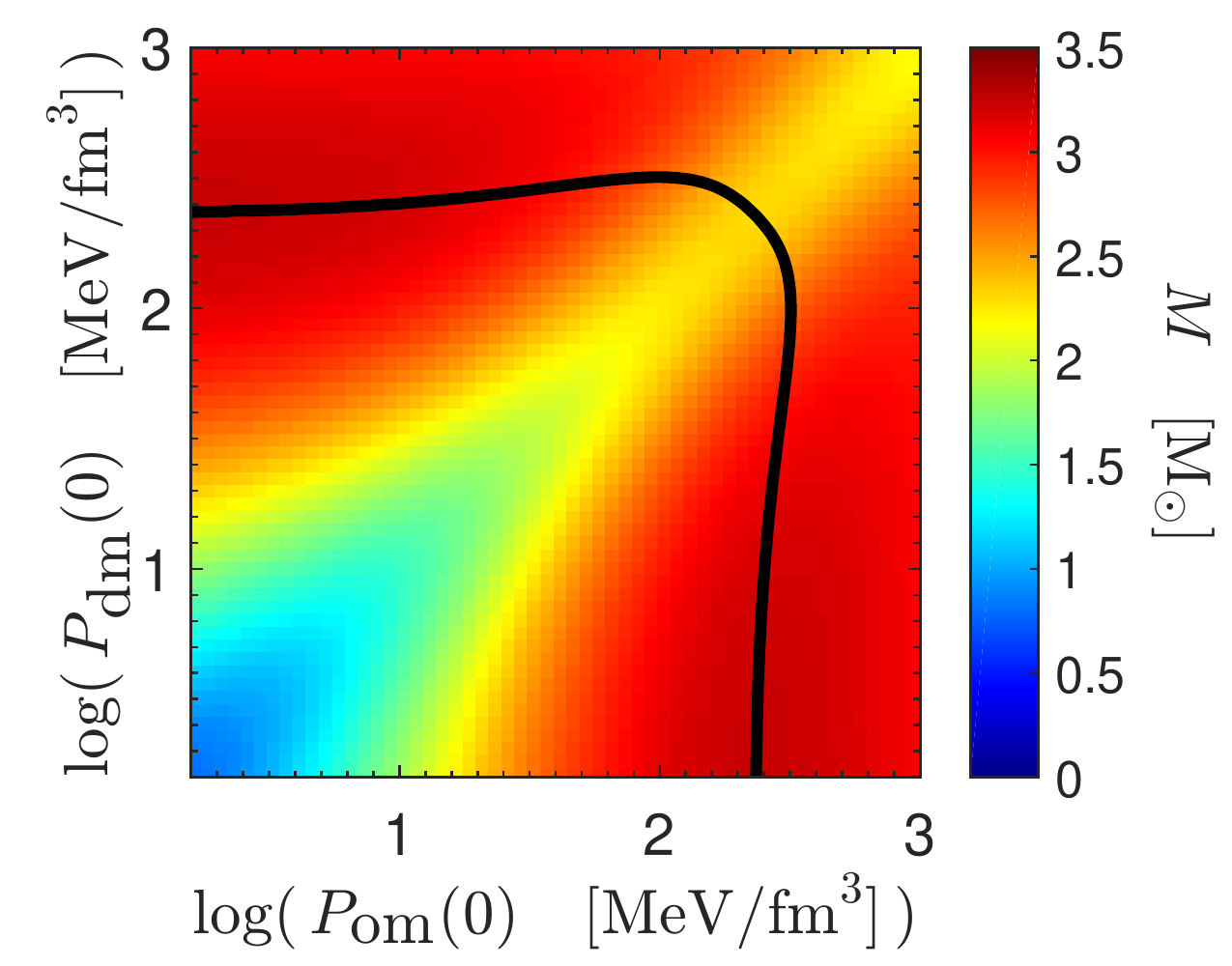}
\caption{Each point in this plot represents a static solution for a dark matter admixed neutron star.  Since the same equation of state is being used for ordinary and dark matter, this plot is symmetric in the two fluids.  The thick black line is the critical curve.  Static solutions below and to the left of the curve are linearly stable, otherwise they are unstable.}
\label{fig:crit curve}
\end{figure}

In Fig.\ \ref{fig:crit curve}, we show the critical curve as the thick black line, as computed using Eq.\ (\ref{stability conditions}).  Each point in Fig.\ \ref{fig:crit curve} represents a static solution.  Since both ordinary and dark matter have the same equation of state, Fig.\ \ref{fig:crit curve} is symmetric in the two fluids, but this does not hold in general \cite{Kain:2021hpk}.  Static solutions enclosed by the critical curve (i.e.\ below and to the left of the curve) are linearly stable; those outside the curve are unstable.  If $P_\text{om}(0)$ or $P_\text{dm}(0)$ is sufficiently small, the other fluid dominates and we effectively have a single-fluid system.  As a consequence, the critical curve reproduces the single-fluid critical pressure in Eq.\ (\ref{1 fluid crit}).  Interestingly, in the upper right corner, there exist linearly stable static solutions that would not be deemed so by a naive single-fluid analysis, which appears to be a general phenomena \cite{Kain:2021hpk}.

A linear stability analysis of dark matter admixed neutron stars, as analyzed using the two methods described above, was given in \cite{Kain:2021hpk}.  Here, we extend this analysis with a study of nonlinear stability by dynamically evolving static solutions.  We have performed a number of evolutions, using both linearly stable and unstable static solutions as initial data.

\begin{figure}
\centering
\includegraphics[width=3.2in]{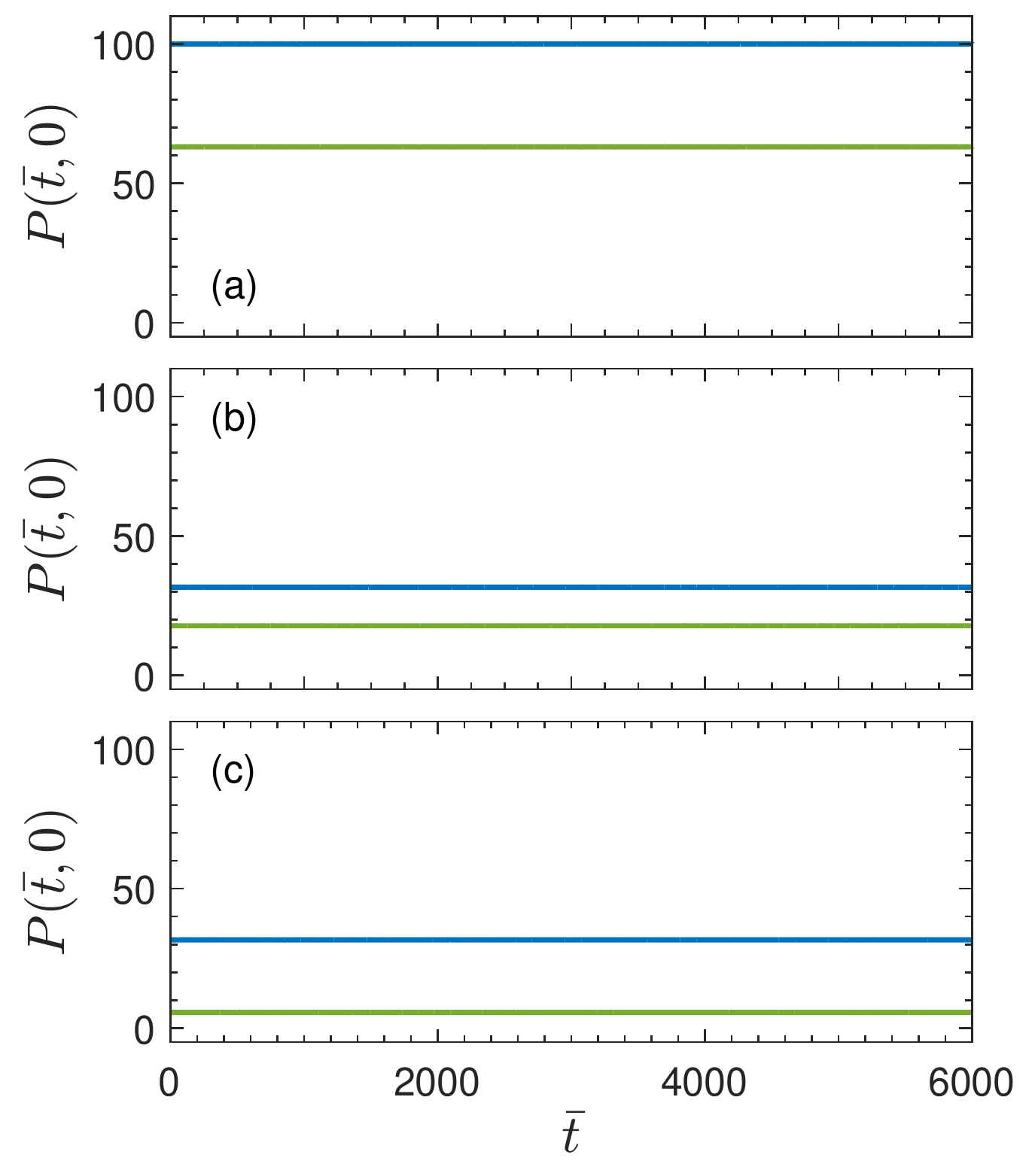}
\caption{The central pressure (in units of MeV/fm$^3$) as a function of time for three evolutions.  In each case, the top blue curve is for ordinary matter and the bottom green curve is for dark matter.  The static solutions used as initial data are defined by $P_\text{om}(0)$, $P_\text{dm}(0) =$ (a) $10^{2}$, $10^{1.8}$, (b) $10^{1.5}$, $10^{1.25}$, and (c) $10^{1.5}$, $10^{0.75}$ MeV/fm$^3$.  That all curves remain remain straight indicates that these three linearly stable static solutions are also nonlinearly stable.}
\label{fig:2 fluid stable}
\end{figure}

We show three evolutions that use a linearly stable static solution as initial data in Fig.\ \ref{fig:2 fluid stable}.  In each case, we plot $P_\text{om}(\bar{t}, 0)$ as the upper blue line and $P_\text{dm}(\bar{t}, 0)$ as the lower green line.  The static solutions used as initial data are given in the caption.  That the curves remain straight indicates that the linearly stable static solutions are also nonlinearly stable.  We find similar results for other linearly stable stars we evolved.  For every unstable static solution we evolved, the system collapsed to a black hole, just as it did in the single-fluid case.

\subsection{Radial oscillation frequencies}
\label{sec:2 fluid freq}

Radial oscillation frequencies for dark matter admixed neutron stars have been computed from pulsation equations in \cite{Leung:2011zz, Leung:2012vea, Kain:2020zjs, Kain:2021hpk}.  In this subsection, we extend the tools available for their study by computing them for the first time from dynamical evolutions.

\begin{figure}
\centering
\includegraphics[width=3.0in]{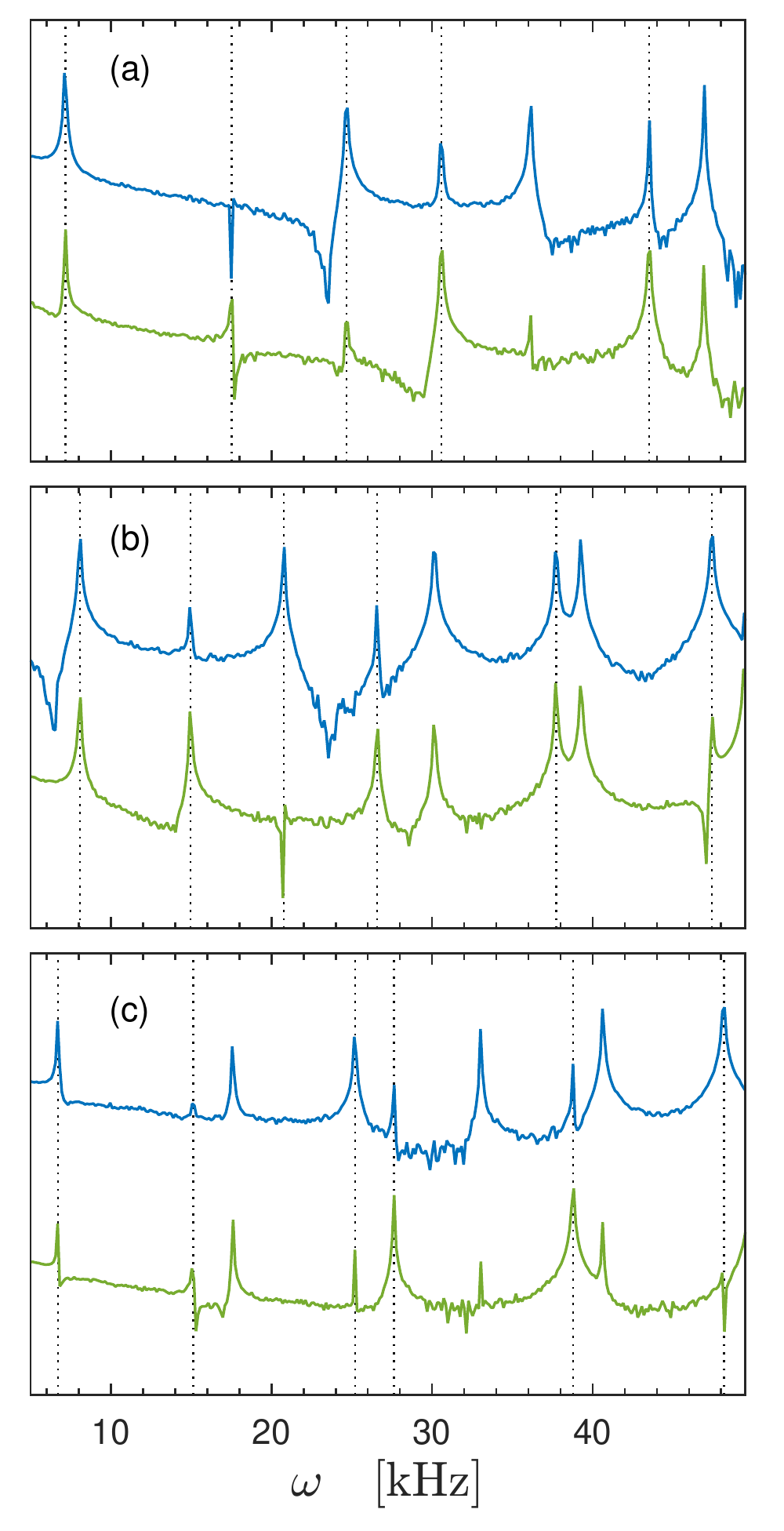}
\caption{The Fourier transform for each of the three evolutions shown in Fig.\ \ref{fig:2 fluid stable} (the units for the vertical axis are arbitrary).  In each case, the upper blue curve is the Fourier transform of ordinary matter and the lower green curve is the Fourier transform of dark matter.  The spikes indicate radial oscillation frequencies.  The dotted vertical lines are independent computations of linear radial oscillation frequencies as computed from pulsation equations.}
\label{fig:2 fluid FT}
\end{figure}

In Fig.\ \ref{fig:2 fluid FT}, we show the Fourier transform of each of the dynamical evolutions shown in Fig.\ \ref{fig:2 fluid stable}.  For each plot in Fig.\ \ref{fig:2 fluid FT}, the top blue curve gives the Fourier transform of $P_\text{om}(\bar{t},0)$ and the bottom green curve gives the Fourier transform of $P_\text{dm}(\bar{t},0)$.  The spikes along the curves correspond to oscillation frequencies.  The dotted vertical lines are the radial oscillation frequencies as computed from pulsation equations, using the methods of \cite{Kain:2020zjs, Kain:2021hpk}.  We can see that the Fourier spectrum and the dotted vertical lines line up well, which is a nontrivial check on the numerical methods developed here and in \cite{Kain:2020zjs}. 

The excitation of a particular radial oscillation mode in a dynamical evolution depends on the perturbations present in the initial data.  Since we are using discretization error as our perturbation, we cannot be sure that all radial oscillation modes will be excited.  However, at least for the equation of state that we are using, this is the case in both the single-fluid and two-fluid cases, since we find spikes in the Fourier spectrum for each of the dotted vertical lines in Figs.\ \ref{fig:1 fluid FT} and \ref{fig:2 fluid FT}.

Dynamical evolutions include both linear and nonlinear oscillations and it may not be clear from the Fourier spectrum which spikes correspond to linear oscillations.  By comparing the Fourier spectrum with the solutions to the pulsation equation, we can determine if a spike has a linear origin.  All three plots in Fig.\ \ref{fig:2 fluid FT} contain spikes in both curves at the same frequency, but which do not have a dotted vertical line running through them.  In these cases, we have been unable to find the respective frequency as a solution to the pulsation equation and we conclude that it is a nonlinear oscillation.


\subsection{Dynamical formation}

In this subsection, we present a dynamical formation process for dark matter admixed neutron stars.  Our model, which is similar to the model used for the formation of fermion-boson stars in \cite{DiGiovanni:2020frc, DiGiovanni:2021vlu}, assumes a preexisting neutron star enveloped by a cloud of dark matter.  The neutron star is described by a single-fluid static solution. The dark matter cloud is described by a Gaussian,
\begin{equation} \label{P Gauss}
P_\text{dm}(0,\bar{r}) = A e^{-2(\bar{r}/\bar{s})^2},
\end{equation}
with a dark matter fluid velocity that is initially zero everywhere, $v_\text{dm}(0,r) = 0$.

\begin{figure}
\centering
\includegraphics[width=2.75in]{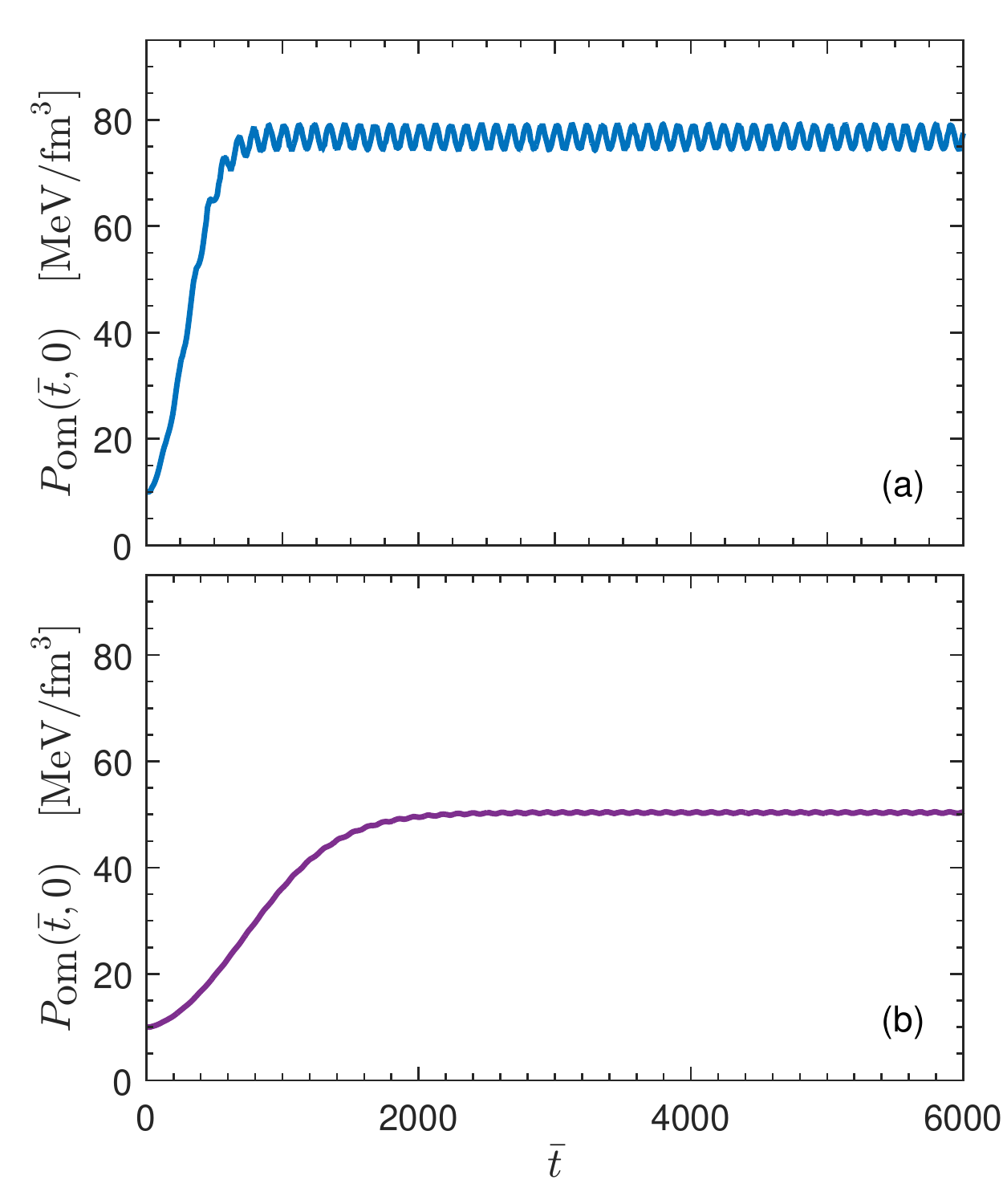}
\caption{Two evolutions are shown for a dynamical formation process.  In both plots, the static solution used for the preexisting neutron star is defined by $P_\text{om}(0) = 10$ MeV/fm$^3$.  The initial dark matter cloud follows from Eq.\ (\ref{P Gauss}) with parameters (a) $\bar{s} = 35$ and $A = 10^{-5}$ MeV/fm$^3$ and (b) $\bar{s} = 70$ and $A = 10^{-7}$ MeV/fm$^3$.  As time moves forward, the system equilibrates and a stable dark matter admixed neutron star is formed.}
\label{fig:formation}
\end{figure}

In Fig.\ \ref{fig:formation}, we show two results.  Both results use the same single-fluid static solution defined by $P_\text{om}(0) = 10$ MeV/fm$^3$ for the preexisting neutron star, which is the same static solution used as initial data for the bottom curve in Fig.\ \ref{fig:1 fluid FT}(a).  For dark matter, Fig.\ \ref{fig:formation}(a) uses $\bar{s} = 35$ and $A = 10^{-5}$ MeV/fm$^3$ and Fig.\ \ref{fig:formation}(b) uses $\bar{s} = 70$ and $A = 10^{-7}$ MeV/fm$^3$.  Figure \ref{fig:formation} plots the central pressure for ordinary matter as a function of time.  We can see that as time moves forward, the central pressure increases.  Eventually, a steady-state is reached, and the star equilibrates to a stable dark matter admixed neutron star.  

\begin{figure*}
\centering
\includegraphics[width=6.75in]{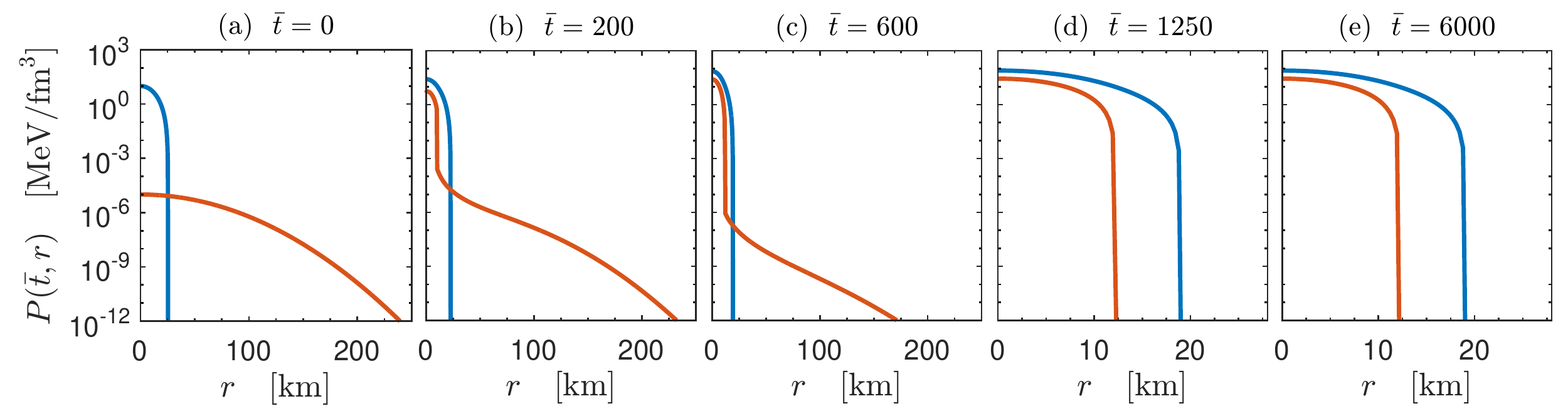}
\caption{Snapshots in time for the same evolution shown in Fig.\ \ref{fig:formation}(a).  In each plot, the blue curve is for ordinary matter and the orange curve is for dark matter.  As time moves forward, dark matter moves inward and the central pressures rise.  At around $\bar{t} = 1250$, a stable dark matter admixed neutron star forms.}
\label{fig:formation time evo}
\end{figure*}

Additional insight can be gained from Fig.\ \ref{fig:formation time evo}, where we show a handful of snapshots for the same evolution shown in Fig.\ \ref{fig:formation}(a).  In each plot, the blue curve plots $P_\text{om}$ and the orange curve plots $P_\text{dm}$.  Figure \ref{fig:formation time evo}(a) shows the starting point, with a cloud of dark matter enveloping a fully formed neutron star.  As time moves forward, dark matter moves inward due to gravitational attraction, and the central pressures rise.  At around $\bar{t} = 1250$, a stable configuration, with a dark matter core, is formed.  For this evolution, the preexisting neutron star has a mass of 2.13 M$_\odot$ and a radius of $25.5$ km.  The final dark matter admixed neutron star has a mass of 2.46 M$_\odot$ and a radius of $19.15$ km.

We have studied a number of evolutions, using various static solutions for the neutron star and various values for the dark matter parameters in Eq.\ (\ref{P Gauss}).  As $s$ or $A$ is increased, more dark matter is included in the system and it becomes possible for the system to collapse to a black hole.  Indeed, it appears to be much easier to get this system to collapse than for the fermion-boson star system \cite{DiGiovanni:2020frc, DiGiovanni:2021vlu}, since fermions do not exhibit gravitational cooling \cite{Seidel:1993zk, DiGiovanni:2018bvo}.  

We end this subsection by stressing that the specific masses and radii that occur in these evolutions should not be taken too seriously, since we are not using a realistic equation of state for ordinary matter, nor a realistic profile for the dark matter cloud.  The purpose of these evolutions is to show that the dynamical formation of dark matter admixed neutron stars is possible.


\section{Conclusion}
\label{sec:conclusion}

The elusive nature of dark matter makes indirect methods for measuring its properties pertinent.  Neutron stars could contain a mixture of ordinary nuclear matter and dark matter, such that dark matter affects bulk properties of the star.  Measurements of neutron stars could then indirectly probe dark matter.  We presented dynamical evolutions of dark matter admixed neutron stars with fermionic dark matter.  We derived the equations of motion, in conservation form, for spherically symmetric systems with an arbitrary number of perfect fluids and developed a hydrodynamical code for solving the two-fluid case.

An aim of this work is to showcase various uses of our hydrodynamical code in the study of dark matter admixed neutron stars.  The linear stability of such systems has been studied in the past.  We showed that linearly stable static solutions can be evolved and determined to also be nonlinearly stable.  Radial oscillation frequencies offer an interesting signature for dark matter admixed neutron stars.  Prior to this work, pulsation equations and dynamical solutions had not both been available for the same type of mixed star.  We solved for oscillation frequencies using both methods independently and showed that the two methods agree.  Finally, dynamical evolutions are likely the only theoretical method available for studying dynamical formation processes.  We showed that it is possible for a dark matter admixed neutron star to form from a preexisting neutron star enveloped in a cloud of dark matter.


\acknowledgments

We thank J.\ Nyhan for help during the early stages of this work.  T.\ G.\ was funded by the 2021 Weiss Summer Research Program.  B.\ B. was funded by a J.\ D.\ Power Center Research Associate grant.


\appendix

\section{Pulsation equations}
\label{app:pulsation}

Chandrasekhar was the first to study stellar oscillations when he derived a pulsation equation for a single prefect fluid \cite{Chandrasekhar:1964zz}.  Recently, this was generalized to an arbitrary number of perfect fluids with only gravitational interfluid interactions \cite{Kain:2020zjs}.  Solutions to the pulsation equations include the squared radial oscillation frequencies, which we compared to the frequencies obtained from dynamical solutions in Secs.\ \ref{sec:1 fluid} and \ref{sec:2 fluid freq}.  In this Appendix, we briefly review the pulsation equations.  We refer the reader to \cite{Chandrasekhar:1964zz, Kain:2020zjs} for additional information.

To derive the pulsation equations, all functions are written as time dependent perturbations about their static solutions and the relevant equations are written to linear order in the perturbations.  Defining $\partial_t \xi_x \equiv \alpha_0 u_x^r$ and then 
\begin{equation} 
\zeta_x \equiv \frac{r^2}{\alpha_0} \xi_x,
\end{equation}
where a subscripted 0 in this appendix refers to a static solution, the equations can be combined such that we obtain a system of pulsation equations that depend on $\zeta_x$ and its derivatives and not on any other perturbations.  Writing the time dependence in harmonic form,
\begin{equation}
\zeta_x(t,r) = \zeta_x(r) e^{i\omega t},
\end{equation}
defines the radial oscillation frequency, $\omega$.

Before presenting the pulsation equations, we parametrize the metric in (\ref{metric}) as
\begin{equation}
ds^2 = - H \sigma^2 dt^2 + \frac{dr^2}{H} + r^2d\Omega^2,
\end{equation}
where
\begin{equation}
H \equiv 1 - \frac{2 m}{r}, 
\qquad
\sigma \equiv \frac{\alpha}{\sqrt{H}}
\end{equation}
($m$ is the same metric function defined in (\ref{m def})), because the metric functions $\sigma$ and $m$ are better suited for solving the pulsation equations.  Writing the first two metric equations in (\ref{metric equations}) in terms of $\sigma$ and $m$ gives
\begin{equation}
\partial_r\sigma_0 = \frac{4\pi G r \sigma_0}{H_0} \sum_y ( \rho_{y0} + P_{y0})
\end{equation}
and the second equation in (\ref{TOV}).

The system of pulsation equations is \cite{Kain:2020zjs}
\begin{widetext}
\begin{align}
&\partial_r (\widehat{\Pi} \zeta_x') +(\widehat{Q}_x + \hat{\omega}^2 W_x) \hat{\zeta}_x
+ \widehat{R}
\left[ 
\left(\frac{\rho_{x0} + P_{x0} }{r} - P_{x0}' \right) 
\sum_y
(\rho_{y0} + P_{y0}) \hat{\zeta}_y
+
\frac{r^2 (\rho_{x0} + P_{x0})}{\hat{\sigma}_0^2 H_0}
\sum_y
\eta_y 
\right]
\notag \\
&\qquad = \widehat{S}_x \sum_y (\rho_{y0} + P_{y0}) \left( \hat{\zeta}_y - \hat{\zeta}_x \right)
+ 
\frac{r^2}{\hat{\sigma}_0^2 H_0}  
\widehat{R}^2
(\rho_{x0} + P_{x0}) 
\sum_y
\sum_z
 P_{y0} \gamma_y (\rho_{kz} + P_{z0}) \left( \hat{\zeta}_z - \hat{\zeta}_y \right)
\notag \\
&\qquad\qquad  +
\widehat{R} \gamma_x P_{x0}
\sum_y \left[(\rho_{y0}' + P_{y0}') \left( \hat{\zeta}_y - \hat{\zeta}_x \right)
+ (\rho_{y0} + P_{y0}) \left( \hat{\zeta}_y' - \hat{\zeta}_x' \right) \right],
\label{pulsation}
\end{align}
where a prime denotes an $r$ derivative, where $P_{x0}'$ is given by the bottom equation in (\ref{TOV}), and where
\begingroup
\allowdisplaybreaks
\begin{align}
\widehat{\Pi}_x &= \frac{1}{r^2}P_{x0} \gamma_i \hat{\sigma}^2_0 H_0
\notag \\
W_x &= \frac{1}{r^2 H_0} (\rho_{x0} + P_{x0} )
\notag \\
\widehat{Q}_x &=
-\frac{ \hat{\sigma}^2_0 H_0}{r^2}
\biggl\{
\frac{3}{r} P_{x0}'
+ 
 \biggl[
\frac{8\pi G}{H_0} P_0
(\rho_{x0} + P_{x0}) 
+ G \left(\frac{4\pi r}{H_0} \sum_y\rho_{y0} - \frac{m_0}{r^2 H_0} \right)  \left( \frac{\rho_{x0} + P_{x0} }{r} - P_{x0}'\right)
\biggr] 
\biggr\}
\notag \\
\widehat{R} &= 4\pi G
\frac{\hat{\sigma}_0^2}{r}
\notag \\
\widehat{S}_x &= \widehat{R}
\biggr\{
(\gamma_x-1)
P_{x0}' 
+ \gamma_x' P_{x0} +
\gamma_x  P_{x0} \bigg[ \frac{8\pi G r}{H_0}\sum_y (\rho_{y0} + P_{y0} )  - \frac{1}{r}\biggr]
\biggr\}
\notag \\
\gamma_x &= \left(1 + \frac{\rho_{x0}}{p_{x0}}\right) \frac{\partial p_{x0}}{\partial \rho_{x0}}.
\label{pulsation defs}
\end{align}
\endgroup
\end{widetext}
Those quantities with a hat have been scaled by powers of $\sigma_0(0)$, the central value of $\sigma_0$, which has the effect of changing the boundary conditions and making the equations easier to solve.  Note that the right hand side of Eq.\ (\ref{pulsation}) vanishes for a single fluid, in which case the left hand side is equivalent to Chandrasekhar's pulsation equation \cite{Chandrasekhar:1964zz}.  Though equivalent, it is not written in an identical form to Chandrasekhar's because, in the presence of multiple fluids, terms cannot cancel and combine in the same way.

For boundary conditions and our method of solution, see \cite{Kain:2020zjs}.  Once the pulsation equations are solved for $\hat{\omega}^2$, and assuming it is positive, the radial oscillation frequency is given by $\omega = \sigma_0(0) \hat{\omega}$.


\section{Spectral decomposition of Jacobian matrix}
\label{app:Jacobian}

In Sec.\ \ref{sec:numerical}, we outlined the numerical methods we used to solve the equations of motion.  These methods included the approximate Riemann solvers HLLE and Roe (see, for example, \cite{RezzollaBook, Neilsen:1999we, Guzman:2012rn}).  Both of these solvers make use of the spectral decomposition of the Jacobian matrix
\begin{equation}
A_x = \frac{\partial \mathbf{f}_x}{\partial \mathbf{u}_x},
\end{equation}
where from (\ref{conservation form 2}),
\begin{equation}
\mathbf{u}_x = 
\begin{pmatrix}
\Pi_x \\ \Phi_x
\end{pmatrix}
\qquad
\mathbf{f}_x = 
\begin{pmatrix}
\frac{1}{2}(\Pi_x - \Phi_x)( 1 + v_x) + P_x \\
\frac{1}{2}(\Pi_x - \Phi_x)( 1 - v_x) - P_x
\end{pmatrix}.
\end{equation}
Specifically, Roe uses the eigenvalues and eigenvectors and HLLE uses just the eigenvalues.

The matrix $A_x$ is $2\times 2$, 
\begin{equation}
A_x = 
\begin{pmatrix}
A_x^{11} & A_x^{12} \\
A_x^{21} & A_x^{22}
\end{pmatrix},
\end{equation}
where \cite{Neilsen:1999we}
\begin{equation}
\begin{split}
A_x^{11} &= \frac{1}{2}( 1+ 2v_x -v_x^2) + (1 - v_x^2) \frac{\partial P_x}{\partial \Pi_x}
\\
A_x^{12}&= -\frac{1}{2} (1 + v_x)^2 + (1 - v_x^2) \frac{\partial P_x}{\partial \Phi_x}
\\
A_x^{21} &= \frac{1}{2}( 1 - v_x)^2 - (1 - v_x^2) \frac{\partial P_x}{\partial \Pi_x}
\\
A_x^{22} &= \frac{1}{2}(- 1+ 2v_x +v_x^2) - (1 - v_x^2) \frac{\partial P_x}{\partial \Phi_x}
\end{split}
\end{equation}
and 
\begin{equation} 
\begin{split}
\frac{\partial P_x}{\partial \Pi_x}
&= \frac{\rho_x - P_x - 2 \Phi_x}{(\Pi_x + \Phi_x - 2\rho_x)
-\frac{\partial \rho_x}{\partial P_x} (\Pi_x + \Phi_x + 2 P_x)}
\\
\frac{\partial P}{\partial \Phi_x}
&= \frac{\rho_x - P_x - 2 \Pi_x}{(\Pi_x + \Phi_x - 2\rho_x)
-\frac{\partial \rho_x}{\partial P_x} (\Pi_x + \Phi_x + 2 P_x)}.
\end{split}
\end{equation}
The derivative $\partial \rho_x / \partial P_x$ is specific to the equation of state.  

The eigenvalues of $A_x$ are \cite{Neilsen:1999we}
\begin{equation}
\lambda_x^\pm = \frac{1}{2} \left[ \text{tr}\, A_x \pm \sqrt{ (\text{tr}\, A_x)^2 - 4 \, \text{det}\, A_x}\right],
\end{equation}
where 
\begin{equation}
\begin{split}
\text{tr}\, A_x &= A_x^{11} + A_x^{22}
\\
\text{det}\, A_x &= A_x^{11}A_x^{22} - A_x^{12}A_x^{21},
\end{split}
\end{equation}
and the eigenvectors are
\begin{equation}
\mathbf{v}_x^\pm = 
\begin{pmatrix}
1 \\
(\lambda_x^\pm - A_x^{11})/A_x^{12}
\end{pmatrix}.
\end{equation}


\section{Code tests}
\label{app:code tests}

In this appendix, we show that our code is second-order convergent.  The middle equation in (\ref{metric eqs}) is a constraint equation for the metric function $a$ and is what we used in our code.  Since we did not use the bottom equation in (\ref{metric eqs}), which is an evolution equation for $a$, it is available for code testing.  We define the constraint
\begin{equation} \label{constraint}
c_a(t,r) \equiv a_\text{code}(t,r) - a_\text{evo}(t,r),
\end{equation}
where $a_\text{code}$ is the value of $a$ used by our code and $a_\text{evo}$ is the value of $a$ obtained from the evolution equation.  

\begin{figure}
\centering
\includegraphics[width=3in]{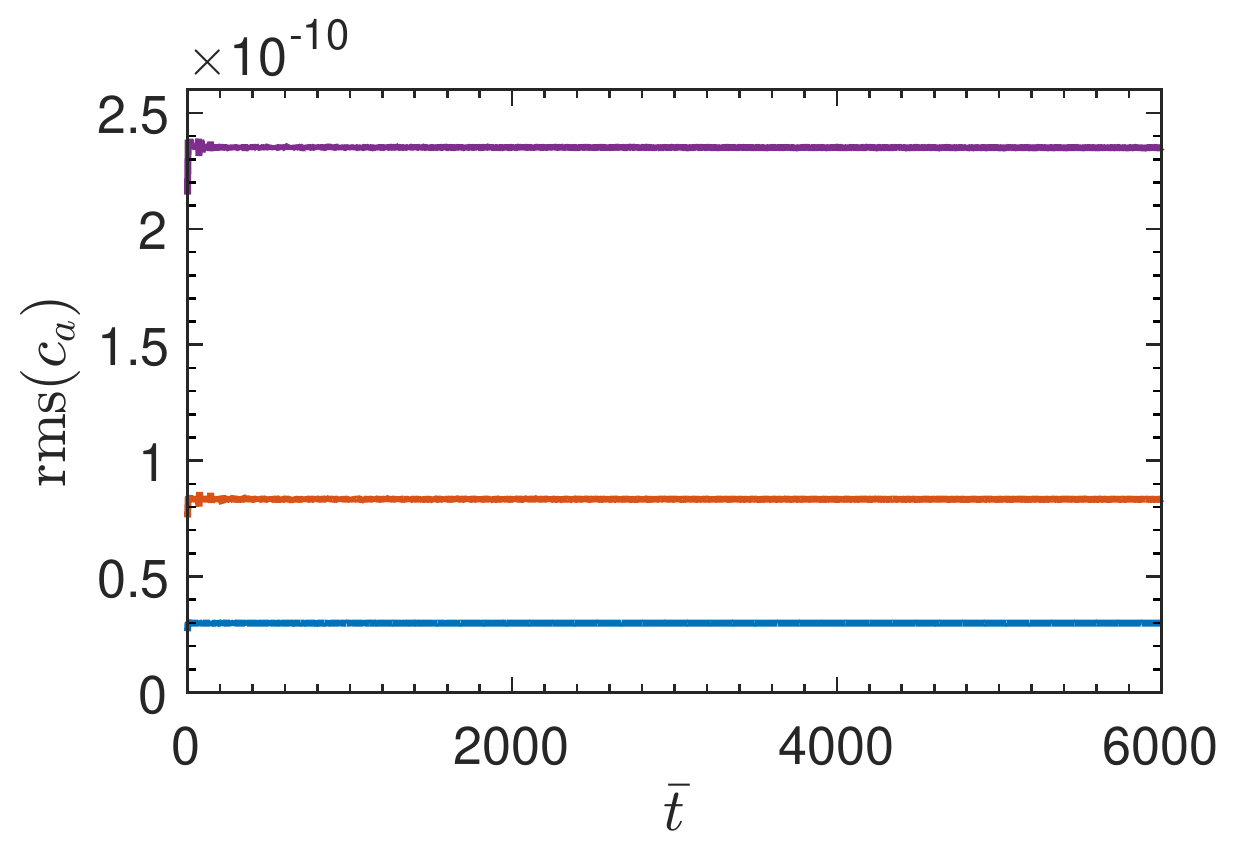}
\caption{The root-mean-square (rms) across the computational grid of the constraint in Eq.\ (\ref{constraint}).  Each curve corresponds to evolving the same initial data, but with a different uniform grid spacing.  From top to bottom, the grid spacings used are $\Delta \bar{r} = 0.02$ (purple), $0.02/\sqrt{2}$ (orange), and 0.01 (blue).  That the curves drop by (at least) a factor of 2 when the grid spacing drops by a factor of $\sqrt{2}$ indicates second order convergence.}
\label{fig:code tests}
\end{figure}

Figure \ref{fig:code tests} shows the root-mean-square (rms) of $c_a$ across the computational grid for a two-fluid dynamical evolution that uses a static solution for initial data defined by $P_\text{om}(0)$, $P_\text{dm}(0) = 10^{1.5}$, $10^{1.25}$ MeV/fm$^3$.  The three curves are for three different grid spacings:\ $\Delta \bar{r} = 0.02$ (top, purple), $0.02/\sqrt{2}$ (middle, orange), and 0.01 (bottom, blue) (for $\Delta \bar{r} = 0.01$, this is the same evolution shown in Figs.\ \ref{fig:2 fluid stable}(b) and \ref{fig:2 fluid FT}(b)).  That the results in Fig.\ \ref{fig:code tests} are small indicates that the constraint $c_a = 0$ is obeyed and that the results drop by (at least) a factor of 2 when the grid spacing drops by a factor of $\sqrt{2}$ indicates second order convergence.

\begin{thebibliography}{79}%
\makeatletter
\providecommand \@ifxundefined [1]{%
 \@ifx{#1\undefined}
}%
\providecommand \@ifnum [1]{%
 \ifnum #1\expandafter \@firstoftwo
 \else \expandafter \@secondoftwo
 \fi
}%
\providecommand \@ifx [1]{%
 \ifx #1\expandafter \@firstoftwo
 \else \expandafter \@secondoftwo
 \fi
}%
\providecommand \natexlab [1]{#1}%
\providecommand \enquote  [1]{``#1''}%
\providecommand \bibnamefont  [1]{#1}%
\providecommand \bibfnamefont [1]{#1}%
\providecommand \citenamefont [1]{#1}%
\providecommand \href@noop [0]{\@secondoftwo}%
\providecommand \href [0]{\begingroup \@sanitize@url \@href}%
\providecommand \@href[1]{\@@startlink{#1}\@@href}%
\providecommand \@@href[1]{\endgroup#1\@@endlink}%
\providecommand \@sanitize@url [0]{\catcode `\\12\catcode `\$12\catcode
  `\&12\catcode `\#12\catcode `\^12\catcode `\_12\catcode `\%12\relax}%
\providecommand \@@startlink[1]{}%
\providecommand \@@endlink[0]{}%
\providecommand \url  [0]{\begingroup\@sanitize@url \@url }%
\providecommand \@url [1]{\endgroup\@href {#1}{\urlprefix }}%
\providecommand \urlprefix  [0]{URL }%
\providecommand \Eprint [0]{\href }%
\providecommand \doibase [0]{https://doi.org/}%
\providecommand \selectlanguage [0]{\@gobble}%
\providecommand \bibinfo  [0]{\@secondoftwo}%
\providecommand \bibfield  [0]{\@secondoftwo}%
\providecommand \translation [1]{[#1]}%
\providecommand \BibitemOpen [0]{}%
\providecommand \bibitemStop [0]{}%
\providecommand \bibitemNoStop [0]{.\EOS\space}%
\providecommand \EOS [0]{\spacefactor3000\relax}%
\providecommand \BibitemShut  [1]{\csname bibitem#1\endcsname}%
\let\auto@bib@innerbib\@empty
\bibitem [{\citenamefont {Henriques}\ \emph {et~al.}(1989)\citenamefont
  {Henriques}, \citenamefont {Liddle},\ and\ \citenamefont
  {Moorhouse}}]{Henriques:1989ar}%
  \BibitemOpen
  \bibfield  {author} {\bibinfo {author} {\bibfnamefont {A.~B.}\ \bibnamefont
  {Henriques}}, \bibinfo {author} {\bibfnamefont {A.~R.}\ \bibnamefont
  {Liddle}},\ and\ \bibinfo {author} {\bibfnamefont {R.~G.}\ \bibnamefont
  {Moorhouse}},\ }\bibfield  {title} {\bibinfo {title} {{Combined Boson-Fermion
  Stars}},\ }\href {https://doi.org/10.1016/0370-2693(89)90623-0} {\bibfield
  {journal} {\bibinfo  {journal} {Phys. Lett. B}\ }\textbf {\bibinfo {volume}
  {233}},\ \bibinfo {pages} {99} (\bibinfo {year} {1989})}\BibitemShut
  {NoStop}%
\bibitem [{\citenamefont {Henriques}\ \emph
  {et~al.}(1990{\natexlab{a}})\citenamefont {Henriques}, \citenamefont
  {Liddle},\ and\ \citenamefont {Moorhouse}}]{Henriques:1989ez}%
  \BibitemOpen
  \bibfield  {author} {\bibinfo {author} {\bibfnamefont {A.~B.}\ \bibnamefont
  {Henriques}}, \bibinfo {author} {\bibfnamefont {A.~R.}\ \bibnamefont
  {Liddle}},\ and\ \bibinfo {author} {\bibfnamefont {R.~G.}\ \bibnamefont
  {Moorhouse}},\ }\bibfield  {title} {\bibinfo {title} {{Combined Boson-Fermion
  Stars: Configurations and Stability}},\ }\href
  {https://doi.org/10.1016/0550-3213(90)90514-E} {\bibfield  {journal}
  {\bibinfo  {journal} {Nucl. Phys. B}\ }\textbf {\bibinfo {volume} {337}},\
  \bibinfo {pages} {737} (\bibinfo {year} {1990}{\natexlab{a}})}\BibitemShut
  {NoStop}%
\bibitem [{\citenamefont {Henriques}\ \emph
  {et~al.}(1990{\natexlab{b}})\citenamefont {Henriques}, \citenamefont
  {Liddle},\ and\ \citenamefont {Moorhouse}}]{Henriques:1990xg}%
  \BibitemOpen
  \bibfield  {author} {\bibinfo {author} {\bibfnamefont {A.~B.}\ \bibnamefont
  {Henriques}}, \bibinfo {author} {\bibfnamefont {A.~R.}\ \bibnamefont
  {Liddle}},\ and\ \bibinfo {author} {\bibfnamefont {R.~G.}\ \bibnamefont
  {Moorhouse}},\ }\bibfield  {title} {\bibinfo {title} {{Stability of
  boson-fermion stars}},\ }\href {https://doi.org/10.1016/0370-2693(90)90789-9}
  {\bibfield  {journal} {\bibinfo  {journal} {Phys. Lett. B}\ }\textbf
  {\bibinfo {volume} {251}},\ \bibinfo {pages} {511} (\bibinfo {year}
  {1990}{\natexlab{b}})}\BibitemShut {NoStop}%
\bibitem [{\citenamefont {de~Sousa}\ and\ \citenamefont
  {Tomazelli}(1998)}]{deSousa:1995ye}%
  \BibitemOpen
  \bibfield  {author} {\bibinfo {author} {\bibfnamefont {C.~M.~G.}\
  \bibnamefont {de~Sousa}}\ and\ \bibinfo {author} {\bibfnamefont {J.~L.}\
  \bibnamefont {Tomazelli}},\ }\bibfield  {title} {\bibinfo {title} {{A Model
  for stars of interacting bosons and fermions}},\ }\href
  {https://doi.org/10.1103/PhysRevD.58.123003} {\bibfield  {journal} {\bibinfo
  {journal} {Phys. Rev. D}\ }\textbf {\bibinfo {volume} {58}},\ \bibinfo
  {pages} {123003} (\bibinfo {year} {1998})},\ \Eprint
  {https://arxiv.org/abs/gr-qc/9507043} {arXiv:gr-qc/9507043} \BibitemShut
  {NoStop}%
\bibitem [{\citenamefont {Pisano}\ and\ \citenamefont
  {Tomazelli}(1996)}]{Pisano:1995yk}%
  \BibitemOpen
  \bibfield  {author} {\bibinfo {author} {\bibfnamefont {F.}~\bibnamefont
  {Pisano}}\ and\ \bibinfo {author} {\bibfnamefont {J.~L.}\ \bibnamefont
  {Tomazelli}},\ }\bibfield  {title} {\bibinfo {title} {{Stars of WIMPs}},\
  }\href {https://doi.org/10.1142/S0217732396000667} {\bibfield  {journal}
  {\bibinfo  {journal} {Mod. Phys. Lett. A}\ }\textbf {\bibinfo {volume}
  {11}},\ \bibinfo {pages} {647} (\bibinfo {year} {1996})},\ \Eprint
  {https://arxiv.org/abs/gr-qc/9509022} {arXiv:gr-qc/9509022} \BibitemShut
  {NoStop}%
\bibitem [{\citenamefont {Sakamoto}\ and\ \citenamefont
  {Shiraishi}(1998)}]{Sakamoto:1998aj}%
  \BibitemOpen
  \bibfield  {author} {\bibinfo {author} {\bibfnamefont {K.}~\bibnamefont
  {Sakamoto}}\ and\ \bibinfo {author} {\bibfnamefont {K.}~\bibnamefont
  {Shiraishi}},\ }\bibfield  {title} {\bibinfo {title} {{Exact solutions for
  boson fermion stars in (2+1)-dimensions}},\ }\href
  {https://doi.org/10.1103/PhysRevD.58.124017} {\bibfield  {journal} {\bibinfo
  {journal} {Phys. Rev. D}\ }\textbf {\bibinfo {volume} {58}},\ \bibinfo
  {pages} {124017} (\bibinfo {year} {1998})},\ \Eprint
  {https://arxiv.org/abs/gr-qc/9806040} {arXiv:gr-qc/9806040} \BibitemShut
  {NoStop}%
\bibitem [{\citenamefont {de~Sousa}\ and\ \citenamefont
  {Silveira}(2001)}]{deSousa:2000eq}%
  \BibitemOpen
  \bibfield  {author} {\bibinfo {author} {\bibfnamefont {C.~M.~G.}\
  \bibnamefont {de~Sousa}}\ and\ \bibinfo {author} {\bibfnamefont
  {V.}~\bibnamefont {Silveira}},\ }\bibfield  {title} {\bibinfo {title}
  {{Slowly rotating boson fermion stars}},\ }\href
  {https://doi.org/10.1142/S0218271801001360} {\bibfield  {journal} {\bibinfo
  {journal} {Int. J. Mod. Phys. D}\ }\textbf {\bibinfo {volume} {10}},\
  \bibinfo {pages} {881} (\bibinfo {year} {2001})},\ \Eprint
  {https://arxiv.org/abs/gr-qc/0012020} {arXiv:gr-qc/0012020} \BibitemShut
  {NoStop}%
\bibitem [{\citenamefont {Henriques}\ and\ \citenamefont
  {Mendes}(2005)}]{Henriques:2003yr}%
  \BibitemOpen
  \bibfield  {author} {\bibinfo {author} {\bibfnamefont {A.~B.}\ \bibnamefont
  {Henriques}}\ and\ \bibinfo {author} {\bibfnamefont {L.~E.}\ \bibnamefont
  {Mendes}},\ }\bibfield  {title} {\bibinfo {title} {{Boson - fermion stars:
  Exploring different configurations}},\ }\href
  {https://doi.org/10.1007/s10509-005-4512-1} {\bibfield  {journal} {\bibinfo
  {journal} {Astrophys. Space Sci.}\ }\textbf {\bibinfo {volume} {300}},\
  \bibinfo {pages} {367} (\bibinfo {year} {2005})},\ \Eprint
  {https://arxiv.org/abs/astro-ph/0301015} {arXiv:astro-ph/0301015}
  \BibitemShut {NoStop}%
\bibitem [{\citenamefont {Dzhunushaliev}\ \emph {et~al.}(2011)\citenamefont
  {Dzhunushaliev}, \citenamefont {Folomeev},\ and\ \citenamefont
  {Singleton}}]{Dzhunushaliev:2011ma}%
  \BibitemOpen
  \bibfield  {author} {\bibinfo {author} {\bibfnamefont {V.}~\bibnamefont
  {Dzhunushaliev}}, \bibinfo {author} {\bibfnamefont {V.}~\bibnamefont
  {Folomeev}},\ and\ \bibinfo {author} {\bibfnamefont {D.}~\bibnamefont
  {Singleton}},\ }\bibfield  {title} {\bibinfo {title} {{Chameleon stars}},\
  }\href {https://doi.org/10.1103/PhysRevD.84.084025} {\bibfield  {journal}
  {\bibinfo  {journal} {Phys. Rev. D}\ }\textbf {\bibinfo {volume} {84}},\
  \bibinfo {pages} {084025} (\bibinfo {year} {2011})},\ \Eprint
  {https://arxiv.org/abs/1106.1267} {arXiv:1106.1267 [astro-ph.SR]}
  \BibitemShut {NoStop}%
\bibitem [{\citenamefont {Valdez-Alvarado}\ \emph {et~al.}(2013)\citenamefont
  {Valdez-Alvarado}, \citenamefont {Palenzuela}, \citenamefont {Alic},\ and\
  \citenamefont {Ureña-López}}]{ValdezAlvarado:2012xc}%
  \BibitemOpen
  \bibfield  {author} {\bibinfo {author} {\bibfnamefont {S.}~\bibnamefont
  {Valdez-Alvarado}}, \bibinfo {author} {\bibfnamefont {C.}~\bibnamefont
  {Palenzuela}}, \bibinfo {author} {\bibfnamefont {D.}~\bibnamefont {Alic}},\
  and\ \bibinfo {author} {\bibfnamefont {L.~A.}\ \bibnamefont
  {Ureña-López}},\ }\bibfield  {title} {\bibinfo {title} {{Dynamical
  evolution of fermion-boson stars}},\ }\href
  {https://doi.org/10.1103/PhysRevD.87.084040} {\bibfield  {journal} {\bibinfo
  {journal} {Phys. Rev. D}\ }\textbf {\bibinfo {volume} {87}},\ \bibinfo
  {pages} {084040} (\bibinfo {year} {2013})},\ \Eprint
  {https://arxiv.org/abs/1210.2299} {arXiv:1210.2299 [gr-qc]} \BibitemShut
  {NoStop}%
\bibitem [{\citenamefont {Brito}\ \emph {et~al.}(2015)\citenamefont {Brito},
  \citenamefont {Cardoso},\ and\ \citenamefont {Okawa}}]{Brito:2015yga}%
  \BibitemOpen
  \bibfield  {author} {\bibinfo {author} {\bibfnamefont {R.}~\bibnamefont
  {Brito}}, \bibinfo {author} {\bibfnamefont {V.}~\bibnamefont {Cardoso}},\
  and\ \bibinfo {author} {\bibfnamefont {H.}~\bibnamefont {Okawa}},\ }\bibfield
   {title} {\bibinfo {title} {{Accretion of dark matter by stars}},\ }\href
  {https://doi.org/10.1103/PhysRevLett.115.111301} {\bibfield  {journal}
  {\bibinfo  {journal} {Phys. Rev. Lett.}\ }\textbf {\bibinfo {volume} {115}},\
  \bibinfo {pages} {111301} (\bibinfo {year} {2015})},\ \Eprint
  {https://arxiv.org/abs/1508.04773} {arXiv:1508.04773 [gr-qc]} \BibitemShut
  {NoStop}%
\bibitem [{\citenamefont {Brito}\ \emph {et~al.}(2016)\citenamefont {Brito},
  \citenamefont {Cardoso}, \citenamefont {Macedo}, \citenamefont {Okawa},\ and\
  \citenamefont {Palenzuela}}]{Brito:2015yfh}%
  \BibitemOpen
  \bibfield  {author} {\bibinfo {author} {\bibfnamefont {R.}~\bibnamefont
  {Brito}}, \bibinfo {author} {\bibfnamefont {V.}~\bibnamefont {Cardoso}},
  \bibinfo {author} {\bibfnamefont {C.~F.~B.}\ \bibnamefont {Macedo}}, \bibinfo
  {author} {\bibfnamefont {H.}~\bibnamefont {Okawa}},\ and\ \bibinfo {author}
  {\bibfnamefont {C.}~\bibnamefont {Palenzuela}},\ }\bibfield  {title}
  {\bibinfo {title} {{Interaction between bosonic dark matter and stars}},\
  }\href {https://doi.org/10.1103/PhysRevD.93.044045} {\bibfield  {journal}
  {\bibinfo  {journal} {Phys. Rev. D}\ }\textbf {\bibinfo {volume} {93}},\
  \bibinfo {pages} {044045} (\bibinfo {year} {2016})},\ \Eprint
  {https://arxiv.org/abs/1512.00466} {arXiv:1512.00466 [astro-ph.SR]}
  \BibitemShut {NoStop}%
\bibitem [{\citenamefont {Bezares}\ \emph {et~al.}(2019)\citenamefont
  {Bezares}, \citenamefont {Vigan\`o},\ and\ \citenamefont
  {Palenzuela}}]{Bezares:2019jcb}%
  \BibitemOpen
  \bibfield  {author} {\bibinfo {author} {\bibfnamefont {M.}~\bibnamefont
  {Bezares}}, \bibinfo {author} {\bibfnamefont {D.}~\bibnamefont {Vigan\`o}},\
  and\ \bibinfo {author} {\bibfnamefont {C.}~\bibnamefont {Palenzuela}},\
  }\bibfield  {title} {\bibinfo {title} {{Gravitational wave signatures of dark
  matter cores in binary neutron star mergers by using numerical
  simulations}},\ }\href {https://doi.org/10.1103/PhysRevD.100.044049}
  {\bibfield  {journal} {\bibinfo  {journal} {Phys. Rev. D}\ }\textbf {\bibinfo
  {volume} {100}},\ \bibinfo {pages} {044049} (\bibinfo {year} {2019})},\
  \Eprint {https://arxiv.org/abs/1905.08551} {arXiv:1905.08551 [gr-qc]}
  \BibitemShut {NoStop}%
\bibitem [{\citenamefont {Di~Giovanni}\ \emph {et~al.}(2020)\citenamefont
  {Di~Giovanni}, \citenamefont {Fakhry}, \citenamefont {Sanchis-Gual},
  \citenamefont {Degollado},\ and\ \citenamefont {Font}}]{DiGiovanni:2020frc}%
  \BibitemOpen
  \bibfield  {author} {\bibinfo {author} {\bibfnamefont {F.}~\bibnamefont
  {Di~Giovanni}}, \bibinfo {author} {\bibfnamefont {S.}~\bibnamefont {Fakhry}},
  \bibinfo {author} {\bibfnamefont {N.}~\bibnamefont {Sanchis-Gual}}, \bibinfo
  {author} {\bibfnamefont {J.~C.}\ \bibnamefont {Degollado}},\ and\ \bibinfo
  {author} {\bibfnamefont {J.~A.}\ \bibnamefont {Font}},\ }\bibfield  {title}
  {\bibinfo {title} {{Dynamical formation and stability of fermion-boson
  stars}},\ }\href {https://doi.org/10.1103/PhysRevD.102.084063} {\bibfield
  {journal} {\bibinfo  {journal} {Phys. Rev. D}\ }\textbf {\bibinfo {volume}
  {102}},\ \bibinfo {pages} {084063} (\bibinfo {year} {2020})},\ \Eprint
  {https://arxiv.org/abs/2006.08583} {arXiv:2006.08583 [gr-qc]} \BibitemShut
  {NoStop}%
\bibitem [{\citenamefont {Valdez-Alvarado}\ \emph {et~al.}(2020)\citenamefont
  {Valdez-Alvarado}, \citenamefont {Becerril},\ and\ \citenamefont {Ure\~na
  L\'opez}}]{Valdez-Alvarado:2020vqa}%
  \BibitemOpen
  \bibfield  {author} {\bibinfo {author} {\bibfnamefont {S.}~\bibnamefont
  {Valdez-Alvarado}}, \bibinfo {author} {\bibfnamefont {R.}~\bibnamefont
  {Becerril}},\ and\ \bibinfo {author} {\bibfnamefont {L.~A.}\ \bibnamefont
  {Ure\~na L\'opez}},\ }\bibfield  {title} {\bibinfo {title} {{Fermion-boson
  stars with a quartic self-interaction in the boson sector}},\ }\href
  {https://doi.org/10.1103/PhysRevD.102.064038} {\bibfield  {journal} {\bibinfo
   {journal} {Phys. Rev. D}\ }\textbf {\bibinfo {volume} {102}},\ \bibinfo
  {pages} {064038} (\bibinfo {year} {2020})},\ \Eprint
  {https://arxiv.org/abs/2001.11009} {arXiv:2001.11009 [gr-qc]} \BibitemShut
  {NoStop}%
\bibitem [{\citenamefont {Di~Giovanni}\ \emph
  {et~al.}(2021{\natexlab{a}})\citenamefont {Di~Giovanni}, \citenamefont
  {Fakhry}, \citenamefont {Sanchis-Gual}, \citenamefont {Degollado},\ and\
  \citenamefont {Font}}]{DiGiovanni:2021vlu}%
  \BibitemOpen
  \bibfield  {author} {\bibinfo {author} {\bibfnamefont {F.}~\bibnamefont
  {Di~Giovanni}}, \bibinfo {author} {\bibfnamefont {S.}~\bibnamefont {Fakhry}},
  \bibinfo {author} {\bibfnamefont {N.}~\bibnamefont {Sanchis-Gual}}, \bibinfo
  {author} {\bibfnamefont {J.~C.}\ \bibnamefont {Degollado}},\ and\ \bibinfo
  {author} {\bibfnamefont {J.~A.}\ \bibnamefont {Font}},\ }\bibfield  {title}
  {\bibinfo {title} {{A stabilization mechanism for excited
  fermion\textendash{}boson stars}},\ }\href
  {https://doi.org/10.1088/1361-6382/ac1b45} {\bibfield  {journal} {\bibinfo
  {journal} {Class. Quant. Grav.}\ }\textbf {\bibinfo {volume} {38}},\ \bibinfo
  {pages} {194001} (\bibinfo {year} {2021}{\natexlab{a}})},\ \Eprint
  {https://arxiv.org/abs/2105.00530} {arXiv:2105.00530 [gr-qc]} \BibitemShut
  {NoStop}%
\bibitem [{\citenamefont {Kain}(2021{\natexlab{a}})}]{Kain:2021bwd}%
  \BibitemOpen
  \bibfield  {author} {\bibinfo {author} {\bibfnamefont {B.}~\bibnamefont
  {Kain}},\ }\bibfield  {title} {\bibinfo {title}
  {{Fermion\textendash{}charged-boson stars}},\ }\href
  {https://doi.org/10.1103/PhysRevD.104.043001} {\bibfield  {journal} {\bibinfo
   {journal} {Phys. Rev. D}\ }\textbf {\bibinfo {volume} {104}},\ \bibinfo
  {pages} {043001} (\bibinfo {year} {2021}{\natexlab{a}})},\ \Eprint
  {https://arxiv.org/abs/2108.01404} {arXiv:2108.01404 [gr-qc]} \BibitemShut
  {NoStop}%
\bibitem [{\citenamefont {Karkevandi}\ \emph {et~al.}(2021)\citenamefont
  {Karkevandi}, \citenamefont {Shakeri}, \citenamefont {Sagun},\ and\
  \citenamefont {Ivanytskyi}}]{Karkevandi:2021ygv}%
  \BibitemOpen
  \bibfield  {author} {\bibinfo {author} {\bibfnamefont {D.~R.}\ \bibnamefont
  {Karkevandi}}, \bibinfo {author} {\bibfnamefont {S.}~\bibnamefont {Shakeri}},
  \bibinfo {author} {\bibfnamefont {V.}~\bibnamefont {Sagun}},\ and\ \bibinfo
  {author} {\bibfnamefont {O.}~\bibnamefont {Ivanytskyi}},\ }\bibfield  {title}
  {\bibinfo {title} {{Bosonic Dark Matter in Neutron Stars and its Effect on
  Gravitational Wave Signal}},\ }\href@noop {} {\  (\bibinfo {year} {2021})},\
  \Eprint {https://arxiv.org/abs/2109.03801} {arXiv:2109.03801 [astro-ph.HE]}
  \BibitemShut {NoStop}%
\bibitem [{\citenamefont {Lee}\ \emph {et~al.}(2021)\citenamefont {Lee},
  \citenamefont {Chu},\ and\ \citenamefont {Lin}}]{Lee:2021yyn}%
  \BibitemOpen
  \bibfield  {author} {\bibinfo {author} {\bibfnamefont {B.~K.~K.}\
  \bibnamefont {Lee}}, \bibinfo {author} {\bibfnamefont {M.-c.}\ \bibnamefont
  {Chu}},\ and\ \bibinfo {author} {\bibfnamefont {L.-M.}\ \bibnamefont {Lin}},\
  }\bibfield  {title} {\bibinfo {title} {{Could the GW190814 Secondary
  Component Be a Bosonic Dark Matter Admixed Compact Star?}},\ }\href
  {https://doi.org/10.3847/1538-4357/ac2735} {\bibfield  {journal} {\bibinfo
  {journal} {Astrophys. J.}\ }\textbf {\bibinfo {volume} {922}},\ \bibinfo
  {pages} {242} (\bibinfo {year} {2021})},\ \Eprint
  {https://arxiv.org/abs/2110.05538} {arXiv:2110.05538 [astro-ph.HE]}
  \BibitemShut {NoStop}%
\bibitem [{\citenamefont {Di~Giovanni}\ \emph
  {et~al.}(2021{\natexlab{b}})\citenamefont {Di~Giovanni}, \citenamefont
  {Sanchis-Gual}, \citenamefont {Cerd\'a-Dur\'an},\ and\ \citenamefont
  {Font}}]{DiGiovanni:2021ejn}%
  \BibitemOpen
  \bibfield  {author} {\bibinfo {author} {\bibfnamefont {F.}~\bibnamefont
  {Di~Giovanni}}, \bibinfo {author} {\bibfnamefont {N.}~\bibnamefont
  {Sanchis-Gual}}, \bibinfo {author} {\bibfnamefont {P.}~\bibnamefont
  {Cerd\'a-Dur\'an}},\ and\ \bibinfo {author} {\bibfnamefont {J.~A.}\
  \bibnamefont {Font}},\ }\bibfield  {title} {\bibinfo {title} {{Can
  fermion-boson stars reconcile multi-messenger observations of compact
  stars?}},\ }\href@noop {} {\  (\bibinfo {year} {2021}{\natexlab{b}})},\
  \Eprint {https://arxiv.org/abs/2110.11997} {arXiv:2110.11997 [gr-qc]}
  \BibitemShut {NoStop}%
\bibitem [{\citenamefont {Sandin}\ and\ \citenamefont
  {Ciarcelluti}(2009)}]{Sandin:2008db}%
  \BibitemOpen
  \bibfield  {author} {\bibinfo {author} {\bibfnamefont {F.}~\bibnamefont
  {Sandin}}\ and\ \bibinfo {author} {\bibfnamefont {P.}~\bibnamefont
  {Ciarcelluti}},\ }\bibfield  {title} {\bibinfo {title} {{Effects of mirror
  dark matter on neutron stars}},\ }\href
  {https://doi.org/10.1016/j.astropartphys.2009.09.005} {\bibfield  {journal}
  {\bibinfo  {journal} {Astropart. Phys.}\ }\textbf {\bibinfo {volume} {32}},\
  \bibinfo {pages} {278} (\bibinfo {year} {2009})},\ \Eprint
  {https://arxiv.org/abs/0809.2942} {arXiv:0809.2942 [astro-ph]} \BibitemShut
  {NoStop}%
\bibitem [{\citenamefont {Ciarcelluti}\ and\ \citenamefont
  {Sandin}(2011)}]{Ciarcelluti:2010ji}%
  \BibitemOpen
  \bibfield  {author} {\bibinfo {author} {\bibfnamefont {P.}~\bibnamefont
  {Ciarcelluti}}\ and\ \bibinfo {author} {\bibfnamefont {F.}~\bibnamefont
  {Sandin}},\ }\bibfield  {title} {\bibinfo {title} {{Have neutron stars a dark
  matter core?}},\ }\href {https://doi.org/10.1016/j.physletb.2010.11.021}
  {\bibfield  {journal} {\bibinfo  {journal} {Phys. Lett. B}\ }\textbf
  {\bibinfo {volume} {695}},\ \bibinfo {pages} {19} (\bibinfo {year} {2011})},\
  \Eprint {https://arxiv.org/abs/1005.0857} {arXiv:1005.0857 [astro-ph.HE]}
  \BibitemShut {NoStop}%
\bibitem [{\citenamefont {Goldman}(2011)}]{Goldman:2011aa}%
  \BibitemOpen
  \bibfield  {author} {\bibinfo {author} {\bibfnamefont {I.}~\bibnamefont
  {Goldman}},\ }\bibfield  {title} {\bibinfo {title} {{Implications of Mirror
  Dark Matter on Neutron Stars}},\ }\href
  {https://doi.org/10.5506/APhysPolB.42.2203} {\bibfield  {journal} {\bibinfo
  {journal} {Acta Phys. Polon. B}\ }\textbf {\bibinfo {volume} {42}},\ \bibinfo
  {pages} {2203} (\bibinfo {year} {2011})},\ \Eprint
  {https://arxiv.org/abs/1112.1505} {arXiv:1112.1505 [astro-ph.CO]}
  \BibitemShut {NoStop}%
\bibitem [{\citenamefont {Leung}\ \emph {et~al.}(2011)\citenamefont {Leung},
  \citenamefont {Chu},\ and\ \citenamefont {Lin}}]{Leung:2011zz}%
  \BibitemOpen
  \bibfield  {author} {\bibinfo {author} {\bibfnamefont {S.}~\bibnamefont
  {Leung}}, \bibinfo {author} {\bibfnamefont {M.}~\bibnamefont {Chu}},\ and\
  \bibinfo {author} {\bibfnamefont {L.}~\bibnamefont {Lin}},\ }\bibfield
  {title} {\bibinfo {title} {{Dark-matter admixed neutron stars}},\ }\href
  {https://doi.org/10.1103/PhysRevD.84.107301} {\bibfield  {journal} {\bibinfo
  {journal} {Phys. Rev. D}\ }\textbf {\bibinfo {volume} {84}},\ \bibinfo
  {pages} {107301} (\bibinfo {year} {2011})},\ \Eprint
  {https://arxiv.org/abs/1111.1787} {arXiv:1111.1787 [astro-ph.CO]}
  \BibitemShut {NoStop}%
\bibitem [{\citenamefont {Leung}\ \emph {et~al.}(2012)\citenamefont {Leung},
  \citenamefont {Chu},\ and\ \citenamefont {Lin}}]{Leung:2012vea}%
  \BibitemOpen
  \bibfield  {author} {\bibinfo {author} {\bibfnamefont {S.}~\bibnamefont
  {Leung}}, \bibinfo {author} {\bibfnamefont {M.}~\bibnamefont {Chu}},\ and\
  \bibinfo {author} {\bibfnamefont {L.}~\bibnamefont {Lin}},\ }\bibfield
  {title} {\bibinfo {title} {{Equilibrium Structure and Radial Oscillations of
  Dark Matter Admixed Neutron Stars}},\ }\href
  {https://doi.org/10.1103/PhysRevD.85.103528} {\bibfield  {journal} {\bibinfo
  {journal} {Phys. Rev. D}\ }\textbf {\bibinfo {volume} {85}},\ \bibinfo
  {pages} {103528} (\bibinfo {year} {2012})},\ \Eprint
  {https://arxiv.org/abs/1205.1909} {arXiv:1205.1909 [astro-ph.CO]}
  \BibitemShut {NoStop}%
\bibitem [{\citenamefont {Li}\ \emph {et~al.}(2012{\natexlab{a}})\citenamefont
  {Li}, \citenamefont {Wang},\ and\ \citenamefont {Cheng}}]{Li_2012}%
  \BibitemOpen
  \bibfield  {author} {\bibinfo {author} {\bibfnamefont {X.~Y.}\ \bibnamefont
  {Li}}, \bibinfo {author} {\bibfnamefont {F.~Y.}\ \bibnamefont {Wang}},\ and\
  \bibinfo {author} {\bibfnamefont {K.~S.}\ \bibnamefont {Cheng}},\ }\bibfield
  {title} {\bibinfo {title} {{Gravitational effects of condensate dark matter
  on compact stellar objects}},\ }\href@noop {} {\bibfield  {journal} {\bibinfo
   {journal} {JCAP}\ }\textbf {\bibinfo {volume} {10}},\ \bibinfo {pages}
  {031}},\ \Eprint {https://arxiv.org/abs/1210.1748} {arXiv:1210.1748
  [astro-ph]} \BibitemShut {NoStop}%
\bibitem [{\citenamefont {Li}\ \emph {et~al.}(2012{\natexlab{b}})\citenamefont
  {Li}, \citenamefont {Huang},\ and\ \citenamefont {Xu}}]{Li:2012ii}%
  \BibitemOpen
  \bibfield  {author} {\bibinfo {author} {\bibfnamefont {A.}~\bibnamefont
  {Li}}, \bibinfo {author} {\bibfnamefont {F.}~\bibnamefont {Huang}},\ and\
  \bibinfo {author} {\bibfnamefont {R.-X.}\ \bibnamefont {Xu}},\ }\bibfield
  {title} {\bibinfo {title} {{Too massive neutron stars: The role of dark
  matter?}},\ }\href {https://doi.org/10.1016/j.astropartphys.2012.07.006}
  {\bibfield  {journal} {\bibinfo  {journal} {Astropart. Phys.}\ }\textbf
  {\bibinfo {volume} {37}},\ \bibinfo {pages} {70} (\bibinfo {year}
  {2012}{\natexlab{b}})},\ \Eprint {https://arxiv.org/abs/1208.3722}
  {arXiv:1208.3722 [astro-ph.SR]} \BibitemShut {NoStop}%
\bibitem [{\citenamefont {Leung}\ \emph {et~al.}(2013)\citenamefont {Leung},
  \citenamefont {Chu}, \citenamefont {Lin},\ and\ \citenamefont
  {Wong}}]{Leung:2013pra}%
  \BibitemOpen
  \bibfield  {author} {\bibinfo {author} {\bibfnamefont {S.-C.}\ \bibnamefont
  {Leung}}, \bibinfo {author} {\bibfnamefont {M.-C.}\ \bibnamefont {Chu}},
  \bibinfo {author} {\bibfnamefont {L.-M.}\ \bibnamefont {Lin}},\ and\ \bibinfo
  {author} {\bibfnamefont {K.-W.}\ \bibnamefont {Wong}},\ }\bibfield  {title}
  {\bibinfo {title} {{Dark-matter admixed white dwarfs}},\ }\href
  {https://doi.org/10.1103/PhysRevD.87.123506} {\bibfield  {journal} {\bibinfo
  {journal} {Phys. Rev. D}\ }\textbf {\bibinfo {volume} {87}},\ \bibinfo
  {pages} {123506} (\bibinfo {year} {2013})},\ \Eprint
  {https://arxiv.org/abs/1305.6142} {arXiv:1305.6142 [astro-ph.CO]}
  \BibitemShut {NoStop}%
\bibitem [{\citenamefont {Goldman}\ \emph {et~al.}(2013)\citenamefont
  {Goldman}, \citenamefont {Mohapatra}, \citenamefont {Nussinov}, \citenamefont
  {Rosenbaum},\ and\ \citenamefont {Teplitz}}]{Goldman:2013qla}%
  \BibitemOpen
  \bibfield  {author} {\bibinfo {author} {\bibfnamefont {I.}~\bibnamefont
  {Goldman}}, \bibinfo {author} {\bibfnamefont {R.}~\bibnamefont {Mohapatra}},
  \bibinfo {author} {\bibfnamefont {S.}~\bibnamefont {Nussinov}}, \bibinfo
  {author} {\bibfnamefont {D.}~\bibnamefont {Rosenbaum}},\ and\ \bibinfo
  {author} {\bibfnamefont {V.}~\bibnamefont {Teplitz}},\ }\bibfield  {title}
  {\bibinfo {title} {{Possible Implications of Asymmetric Fermionic Dark Matter
  for Neutron Stars}},\ }\href {https://doi.org/10.1016/j.physletb.2013.07.017}
  {\bibfield  {journal} {\bibinfo  {journal} {Phys. Lett. B}\ }\textbf
  {\bibinfo {volume} {725}},\ \bibinfo {pages} {200} (\bibinfo {year}
  {2013})},\ \Eprint {https://arxiv.org/abs/1305.6908} {arXiv:1305.6908
  [astro-ph.CO]} \BibitemShut {NoStop}%
\bibitem [{\citenamefont {Xiang}\ \emph {et~al.}(2014)\citenamefont {Xiang},
  \citenamefont {Jiang}, \citenamefont {Zhang},\ and\ \citenamefont
  {Yang}}]{Xiang:2013xwa}%
  \BibitemOpen
  \bibfield  {author} {\bibinfo {author} {\bibfnamefont {Q.-F.}\ \bibnamefont
  {Xiang}}, \bibinfo {author} {\bibfnamefont {W.-Z.}\ \bibnamefont {Jiang}},
  \bibinfo {author} {\bibfnamefont {D.-R.}\ \bibnamefont {Zhang}},\ and\
  \bibinfo {author} {\bibfnamefont {R.-Y.}\ \bibnamefont {Yang}},\ }\bibfield
  {title} {\bibinfo {title} {{Effects of fermionic dark matter on properties of
  neutron stars}},\ }\href {https://doi.org/10.1103/PhysRevC.89.025803}
  {\bibfield  {journal} {\bibinfo  {journal} {Phys. Rev. C}\ }\textbf {\bibinfo
  {volume} {89}},\ \bibinfo {pages} {025803} (\bibinfo {year} {2014})},\
  \Eprint {https://arxiv.org/abs/1305.7354} {arXiv:1305.7354 [astro-ph.SR]}
  \BibitemShut {NoStop}%
\bibitem [{\citenamefont {Tolos}\ and\ \citenamefont
  {Schaffner-Bielich}(2015)}]{Tolos:2015qra}%
  \BibitemOpen
  \bibfield  {author} {\bibinfo {author} {\bibfnamefont {L.}~\bibnamefont
  {Tolos}}\ and\ \bibinfo {author} {\bibfnamefont {J.}~\bibnamefont
  {Schaffner-Bielich}},\ }\bibfield  {title} {\bibinfo {title} {{Dark Compact
  Planets}},\ }\href {https://doi.org/10.1103/PhysRevD.92.123002} {\bibfield
  {journal} {\bibinfo  {journal} {Phys. Rev. D}\ }\textbf {\bibinfo {volume}
  {92}},\ \bibinfo {pages} {123002} (\bibinfo {year} {2015})},\ \Eprint
  {https://arxiv.org/abs/1507.08197} {arXiv:1507.08197 [astro-ph.HE]}
  \BibitemShut {NoStop}%
\bibitem [{\citenamefont {Mukhopadhyay}\ and\ \citenamefont
  {Schaffner-Bielich}(2016)}]{Mukhopadhyay:2015xhs}%
  \BibitemOpen
  \bibfield  {author} {\bibinfo {author} {\bibfnamefont {P.}~\bibnamefont
  {Mukhopadhyay}}\ and\ \bibinfo {author} {\bibfnamefont {J.}~\bibnamefont
  {Schaffner-Bielich}},\ }\bibfield  {title} {\bibinfo {title} {{Quark stars
  admixed with dark matter}},\ }\href
  {https://doi.org/10.1103/PhysRevD.93.083009} {\bibfield  {journal} {\bibinfo
  {journal} {Phys. Rev. D}\ }\textbf {\bibinfo {volume} {93}},\ \bibinfo
  {pages} {083009} (\bibinfo {year} {2016})},\ \Eprint
  {https://arxiv.org/abs/1511.00238} {arXiv:1511.00238 [astro-ph.HE]}
  \BibitemShut {NoStop}%
\bibitem [{\citenamefont {Panotopoulos}\ and\ \citenamefont
  {Lopes}(2017{\natexlab{a}})}]{Panotopoulos:2017pgv}%
  \BibitemOpen
  \bibfield  {author} {\bibinfo {author} {\bibfnamefont {G.}~\bibnamefont
  {Panotopoulos}}\ and\ \bibinfo {author} {\bibfnamefont {I.}~\bibnamefont
  {Lopes}},\ }\bibfield  {title} {\bibinfo {title} {{Gravitational effects of
  condensed dark matter on strange stars}},\ }\href
  {https://doi.org/10.1103/PhysRevD.96.023002} {\bibfield  {journal} {\bibinfo
  {journal} {Phys. Rev. D}\ }\textbf {\bibinfo {volume} {96}},\ \bibinfo
  {pages} {023002} (\bibinfo {year} {2017}{\natexlab{a}})},\ \Eprint
  {https://arxiv.org/abs/1706.07272} {arXiv:1706.07272 [gr-qc]} \BibitemShut
  {NoStop}%
\bibitem [{\citenamefont {Panotopoulos}\ and\ \citenamefont
  {Lopes}(2017{\natexlab{b}})}]{Panotopoulos:2017idn}%
  \BibitemOpen
  \bibfield  {author} {\bibinfo {author} {\bibfnamefont {G.}~\bibnamefont
  {Panotopoulos}}\ and\ \bibinfo {author} {\bibfnamefont {I.}~\bibnamefont
  {Lopes}},\ }\bibfield  {title} {\bibinfo {title} {{Dark matter effect on
  realistic equation of state in neutron stars}},\ }\href
  {https://doi.org/10.1103/PhysRevD.96.083004} {\bibfield  {journal} {\bibinfo
  {journal} {Phys. Rev. D}\ }\textbf {\bibinfo {volume} {96}},\ \bibinfo
  {pages} {083004} (\bibinfo {year} {2017}{\natexlab{b}})},\ \Eprint
  {https://arxiv.org/abs/1709.06312} {arXiv:1709.06312 [hep-ph]} \BibitemShut
  {NoStop}%
\bibitem [{\citenamefont {Gresham}\ and\ \citenamefont
  {Zurek}(2019)}]{Gresham:2018rqo}%
  \BibitemOpen
  \bibfield  {author} {\bibinfo {author} {\bibfnamefont {M.~I.}\ \bibnamefont
  {Gresham}}\ and\ \bibinfo {author} {\bibfnamefont {K.~M.}\ \bibnamefont
  {Zurek}},\ }\bibfield  {title} {\bibinfo {title} {{Asymmetric Dark Stars and
  Neutron Star Stability}},\ }\href
  {https://doi.org/10.1103/PhysRevD.99.083008} {\bibfield  {journal} {\bibinfo
  {journal} {Phys. Rev. D}\ }\textbf {\bibinfo {volume} {99}},\ \bibinfo
  {pages} {083008} (\bibinfo {year} {2019})},\ \Eprint
  {https://arxiv.org/abs/1809.08254} {arXiv:1809.08254 [astro-ph.CO]}
  \BibitemShut {NoStop}%
\bibitem [{\citenamefont {Nelson}\ \emph {et~al.}(2019)\citenamefont {Nelson},
  \citenamefont {Reddy},\ and\ \citenamefont {Zhou}}]{Nelson:2018xtr}%
  \BibitemOpen
  \bibfield  {author} {\bibinfo {author} {\bibfnamefont {A.}~\bibnamefont
  {Nelson}}, \bibinfo {author} {\bibfnamefont {S.}~\bibnamefont {Reddy}},\ and\
  \bibinfo {author} {\bibfnamefont {D.}~\bibnamefont {Zhou}},\ }\bibfield
  {title} {\bibinfo {title} {{Dark halos around neutron stars and gravitational
  waves}},\ }\href {https://doi.org/10.1088/1475-7516/2019/07/012} {\bibfield
  {journal} {\bibinfo  {journal} {JCAP}\ }\textbf {\bibinfo {volume} {07}},\
  \bibinfo {pages} {012}},\ \Eprint {https://arxiv.org/abs/1803.03266}
  {arXiv:1803.03266 [hep-ph]} \BibitemShut {NoStop}%
\bibitem [{\citenamefont {Ellis}\ \emph {et~al.}(2018)\citenamefont {Ellis},
  \citenamefont {H\"utsi}, \citenamefont {Kannike}, \citenamefont {Marzola},
  \citenamefont {Raidal},\ and\ \citenamefont {Vaskonen}}]{Ellis:2018bkr}%
  \BibitemOpen
  \bibfield  {author} {\bibinfo {author} {\bibfnamefont {J.}~\bibnamefont
  {Ellis}}, \bibinfo {author} {\bibfnamefont {G.}~\bibnamefont {H\"utsi}},
  \bibinfo {author} {\bibfnamefont {K.}~\bibnamefont {Kannike}}, \bibinfo
  {author} {\bibfnamefont {L.}~\bibnamefont {Marzola}}, \bibinfo {author}
  {\bibfnamefont {M.}~\bibnamefont {Raidal}},\ and\ \bibinfo {author}
  {\bibfnamefont {V.}~\bibnamefont {Vaskonen}},\ }\bibfield  {title} {\bibinfo
  {title} {{Dark Matter Effects On Neutron Star Properties}},\ }\href
  {https://doi.org/10.1103/PhysRevD.97.123007} {\bibfield  {journal} {\bibinfo
  {journal} {Phys. Rev. D}\ }\textbf {\bibinfo {volume} {97}},\ \bibinfo
  {pages} {123007} (\bibinfo {year} {2018})},\ \Eprint
  {https://arxiv.org/abs/1804.01418} {arXiv:1804.01418 [astro-ph.CO]}
  \BibitemShut {NoStop}%
\bibitem [{\citenamefont {Deliyergiyev}\ \emph {et~al.}(2019)\citenamefont
  {Deliyergiyev}, \citenamefont {Del~Popolo}, \citenamefont {Tolos},
  \citenamefont {Le~Delliou}, \citenamefont {Lee},\ and\ \citenamefont
  {Burgio}}]{Deliyergiyev:2019vti}%
  \BibitemOpen
  \bibfield  {author} {\bibinfo {author} {\bibfnamefont {M.}~\bibnamefont
  {Deliyergiyev}}, \bibinfo {author} {\bibfnamefont {A.}~\bibnamefont
  {Del~Popolo}}, \bibinfo {author} {\bibfnamefont {L.}~\bibnamefont {Tolos}},
  \bibinfo {author} {\bibfnamefont {M.}~\bibnamefont {Le~Delliou}}, \bibinfo
  {author} {\bibfnamefont {X.}~\bibnamefont {Lee}},\ and\ \bibinfo {author}
  {\bibfnamefont {F.}~\bibnamefont {Burgio}},\ }\bibfield  {title} {\bibinfo
  {title} {{Dark compact objects: an extensive overview}},\ }\href
  {https://doi.org/10.1103/PhysRevD.99.063015} {\bibfield  {journal} {\bibinfo
  {journal} {Phys. Rev. D}\ }\textbf {\bibinfo {volume} {99}},\ \bibinfo
  {pages} {063015} (\bibinfo {year} {2019})},\ \Eprint
  {https://arxiv.org/abs/1903.01183} {arXiv:1903.01183 [gr-qc]} \BibitemShut
  {NoStop}%
\bibitem [{\citenamefont {Bhat}\ and\ \citenamefont
  {Paul}(2020)}]{Bhat:2019tnz}%
  \BibitemOpen
  \bibfield  {author} {\bibinfo {author} {\bibfnamefont {S.~A.}\ \bibnamefont
  {Bhat}}\ and\ \bibinfo {author} {\bibfnamefont {A.}~\bibnamefont {Paul}},\
  }\bibfield  {title} {\bibinfo {title} {{Cooling of Dark-Matter Admixed
  Neutron Stars with density-dependent Equation of State}},\ }\href
  {https://doi.org/10.1140/epjc/s10052-020-8072-x} {\bibfield  {journal}
  {\bibinfo  {journal} {Eur. Phys. J. C}\ }\textbf {\bibinfo {volume} {80}},\
  \bibinfo {pages} {544} (\bibinfo {year} {2020})},\ \Eprint
  {https://arxiv.org/abs/1905.12483} {arXiv:1905.12483 [hep-ph]} \BibitemShut
  {NoStop}%
\bibitem [{\citenamefont {Del~Popolo}\ \emph {et~al.}(2020)\citenamefont
  {Del~Popolo}, \citenamefont {Deliyergiyev},\ and\ \citenamefont
  {Le~Delliou}}]{DelPopolo:2020pzh}%
  \BibitemOpen
  \bibfield  {author} {\bibinfo {author} {\bibfnamefont {A.}~\bibnamefont
  {Del~Popolo}}, \bibinfo {author} {\bibfnamefont {M.}~\bibnamefont
  {Deliyergiyev}},\ and\ \bibinfo {author} {\bibfnamefont {M.}~\bibnamefont
  {Le~Delliou}},\ }\bibfield  {title} {\bibinfo {title} {{Solution to the
  hyperon puzzle using dark matter}},\ }\href
  {https://doi.org/10.1016/j.dark.2020.100622} {\bibfield  {journal} {\bibinfo
  {journal} {Phys. Dark Univ.}\ }\textbf {\bibinfo {volume} {30}},\ \bibinfo
  {pages} {100622} (\bibinfo {year} {2020})},\ \Eprint
  {https://arxiv.org/abs/2011.00984} {arXiv:2011.00984 [gr-qc]} \BibitemShut
  {NoStop}%
\bibitem [{\citenamefont {Zhang}\ and\ \citenamefont
  {Lin}(2020)}]{Zhang:2020dfi}%
  \BibitemOpen
  \bibfield  {author} {\bibinfo {author} {\bibfnamefont {K.}~\bibnamefont
  {Zhang}}\ and\ \bibinfo {author} {\bibfnamefont {F.-L.}\ \bibnamefont
  {Lin}},\ }\bibfield  {title} {\bibinfo {title} {{Constraint on hybrid stars
  with gravitational wave events}},\ }\href
  {https://doi.org/10.3390/universe6120231} {\bibfield  {journal} {\bibinfo
  {journal} {Universe}\ }\textbf {\bibinfo {volume} {6}},\ \bibinfo {pages}
  {231} (\bibinfo {year} {2020})},\ \Eprint {https://arxiv.org/abs/2011.05104}
  {arXiv:2011.05104 [astro-ph.HE]} \BibitemShut {NoStop}%
\bibitem [{\citenamefont {Das}\ \emph {et~al.}(2020)\citenamefont {Das},
  \citenamefont {Kumar}, \citenamefont {Kumar}, \citenamefont {Kumar~Biswal},
  \citenamefont {Nakatsukasa}, \citenamefont {Li},\ and\ \citenamefont
  {Patra}}]{Das:2020vng}%
  \BibitemOpen
  \bibfield  {author} {\bibinfo {author} {\bibfnamefont {H.~C.}\ \bibnamefont
  {Das}}, \bibinfo {author} {\bibfnamefont {A.}~\bibnamefont {Kumar}}, \bibinfo
  {author} {\bibfnamefont {B.}~\bibnamefont {Kumar}}, \bibinfo {author}
  {\bibfnamefont {S.}~\bibnamefont {Kumar~Biswal}}, \bibinfo {author}
  {\bibfnamefont {T.}~\bibnamefont {Nakatsukasa}}, \bibinfo {author}
  {\bibfnamefont {A.}~\bibnamefont {Li}},\ and\ \bibinfo {author}
  {\bibfnamefont {S.~K.}\ \bibnamefont {Patra}},\ }\bibfield  {title} {\bibinfo
  {title} {{Effects of dark matter on the nuclear and neutron star matter}},\
  }\href {https://doi.org/10.1093/mnras/staa1435} {\bibfield  {journal}
  {\bibinfo  {journal} {Mon. Not. Roy. Astron. Soc.}\ }\textbf {\bibinfo
  {volume} {495}},\ \bibinfo {pages} {4893} (\bibinfo {year} {2020})},\ \Eprint
  {https://arxiv.org/abs/2002.00594} {arXiv:2002.00594 [nucl-th]} \BibitemShut
  {NoStop}%
\bibitem [{\citenamefont {Kain}(2020)}]{Kain:2020zjs}%
  \BibitemOpen
  \bibfield  {author} {\bibinfo {author} {\bibfnamefont {B.}~\bibnamefont
  {Kain}},\ }\bibfield  {title} {\bibinfo {title} {{Radial oscillations and
  stability of multiple-fluid compact stars}},\ }\href
  {https://doi.org/10.1103/PhysRevD.102.023001} {\bibfield  {journal} {\bibinfo
   {journal} {Phys. Rev. D}\ }\textbf {\bibinfo {volume} {102}},\ \bibinfo
  {pages} {023001} (\bibinfo {year} {2020})},\ \Eprint
  {https://arxiv.org/abs/2007.04311} {arXiv:2007.04311 [gr-qc]} \BibitemShut
  {NoStop}%
\bibitem [{\citenamefont {Kain}(2021{\natexlab{b}})}]{Kain:2021hpk}%
  \BibitemOpen
  \bibfield  {author} {\bibinfo {author} {\bibfnamefont {B.}~\bibnamefont
  {Kain}},\ }\bibfield  {title} {\bibinfo {title} {{Dark matter admixed neutron
  stars}},\ }\href {https://doi.org/10.1103/PhysRevD.103.043009} {\bibfield
  {journal} {\bibinfo  {journal} {Phys. Rev. D}\ }\textbf {\bibinfo {volume}
  {103}},\ \bibinfo {pages} {043009} (\bibinfo {year} {2021}{\natexlab{b}})},\
  \Eprint {https://arxiv.org/abs/2102.08257} {arXiv:2102.08257 [gr-qc]}
  \BibitemShut {NoStop}%
\bibitem [{\citenamefont {Das}\ \emph {et~al.}(2021{\natexlab{a}})\citenamefont
  {Das}, \citenamefont {Kumar}, \citenamefont {Biswal},\ and\ \citenamefont
  {Patra}}]{Das:2021dru}%
  \BibitemOpen
  \bibfield  {author} {\bibinfo {author} {\bibfnamefont {H.~C.}\ \bibnamefont
  {Das}}, \bibinfo {author} {\bibfnamefont {A.}~\bibnamefont {Kumar}}, \bibinfo
  {author} {\bibfnamefont {S.~K.}\ \bibnamefont {Biswal}},\ and\ \bibinfo
  {author} {\bibfnamefont {S.~K.}\ \bibnamefont {Patra}},\ }\bibfield  {title}
  {\bibinfo {title} {{Impacts of dark matter on the f-mode oscillation of
  hyperon star}},\ }\href {https://doi.org/10.1103/PhysRevD.104.123006}
  {\bibfield  {journal} {\bibinfo  {journal} {Phys. Rev. D}\ }\textbf {\bibinfo
  {volume} {104}},\ \bibinfo {pages} {123006} (\bibinfo {year}
  {2021}{\natexlab{a}})},\ \Eprint {https://arxiv.org/abs/2109.01851}
  {arXiv:2109.01851 [nucl-th]} \BibitemShut {NoStop}%
\bibitem [{\citenamefont {Das}\ \emph {et~al.}(2021{\natexlab{b}})\citenamefont
  {Das}, \citenamefont {Kumar},\ and\ \citenamefont {Patra}}]{Das:2021yny}%
  \BibitemOpen
  \bibfield  {author} {\bibinfo {author} {\bibfnamefont {H.~C.}\ \bibnamefont
  {Das}}, \bibinfo {author} {\bibfnamefont {A.}~\bibnamefont {Kumar}},\ and\
  \bibinfo {author} {\bibfnamefont {S.~K.}\ \bibnamefont {Patra}},\ }\bibfield
  {title} {\bibinfo {title} {{Dark matter admixed neutron star as a possible
  compact component in the GW190814 merger event}},\ }\href
  {https://doi.org/10.1103/PhysRevD.104.063028} {\bibfield  {journal} {\bibinfo
   {journal} {Phys. Rev. D}\ }\textbf {\bibinfo {volume} {104}},\ \bibinfo
  {pages} {063028} (\bibinfo {year} {2021}{\natexlab{b}})},\ \Eprint
  {https://arxiv.org/abs/2109.01853} {arXiv:2109.01853 [astro-ph.HE]}
  \BibitemShut {NoStop}%
\bibitem [{\citenamefont {Sen}\ and\ \citenamefont {Guha}(2021)}]{Sen:2021wev}%
  \BibitemOpen
  \bibfield  {author} {\bibinfo {author} {\bibfnamefont {D.}~\bibnamefont
  {Sen}}\ and\ \bibinfo {author} {\bibfnamefont {A.}~\bibnamefont {Guha}},\
  }\bibfield  {title} {\bibinfo {title} {{Implications of feebly interacting
  dark sector on neutron star properties and constraints from GW170817}},\
  }\href {https://doi.org/10.1093/mnras/stab1056} {\bibfield  {journal}
  {\bibinfo  {journal} {Mon. Not. Roy. Astron. Soc.}\ }\textbf {\bibinfo
  {volume} {504}},\ \bibinfo {pages} {3354} (\bibinfo {year} {2021})},\ \Eprint
  {https://arxiv.org/abs/2104.06141} {arXiv:2104.06141 [hep-ph]} \BibitemShut
  {NoStop}%
\bibitem [{\citenamefont {Das}\ \emph {et~al.}(2021{\natexlab{c}})\citenamefont
  {Das}, \citenamefont {Kumar},\ and\ \citenamefont {Patra}}]{Das:2021wku}%
  \BibitemOpen
  \bibfield  {author} {\bibinfo {author} {\bibfnamefont {H.~C.}\ \bibnamefont
  {Das}}, \bibinfo {author} {\bibfnamefont {A.}~\bibnamefont {Kumar}},\ and\
  \bibinfo {author} {\bibfnamefont {S.~K.}\ \bibnamefont {Patra}},\ }\bibfield
  {title} {\bibinfo {title} {{Effects of dark matter on the in-spiral
  properties of the binary neutron stars}},\ }\href
  {https://doi.org/10.1093/mnras/stab2387} {\bibfield  {journal} {\bibinfo
  {journal} {Mon. Not. Roy. Astron. Soc.}\ }\textbf {\bibinfo {volume} {507}},\
  \bibinfo {pages} {4053} (\bibinfo {year} {2021}{\natexlab{c}})},\ \Eprint
  {https://arxiv.org/abs/2104.01815} {arXiv:2104.01815 [astro-ph.HE]}
  \BibitemShut {NoStop}%
\bibitem [{\citenamefont {Das}\ \emph {et~al.}(2021{\natexlab{d}})\citenamefont
  {Das}, \citenamefont {Kumar}, \citenamefont {Kumar}, \citenamefont {Biswal},\
  and\ \citenamefont {Patra}}]{Das:2020ptd}%
  \BibitemOpen
  \bibfield  {author} {\bibinfo {author} {\bibfnamefont {H.~C.}\ \bibnamefont
  {Das}}, \bibinfo {author} {\bibfnamefont {A.}~\bibnamefont {Kumar}}, \bibinfo
  {author} {\bibfnamefont {B.}~\bibnamefont {Kumar}}, \bibinfo {author}
  {\bibfnamefont {S.~K.}\ \bibnamefont {Biswal}},\ and\ \bibinfo {author}
  {\bibfnamefont {S.~K.}\ \bibnamefont {Patra}},\ }\bibfield  {title} {\bibinfo
  {title} {{Impacts of dark matter on the curvature of the neutron star}},\
  }\href {https://doi.org/10.1088/1475-7516/2021/01/007} {\bibfield  {journal}
  {\bibinfo  {journal} {JCAP}\ }\textbf {\bibinfo {volume} {01}},\ \bibinfo
  {pages} {007}},\ \Eprint {https://arxiv.org/abs/2007.05382} {arXiv:2007.05382
  [nucl-th]} \BibitemShut {NoStop}%
\bibitem [{\citenamefont {Jim\'enez}\ and\ \citenamefont
  {Fraga}(2021)}]{Jimenez:2021nmr}%
  \BibitemOpen
  \bibfield  {author} {\bibinfo {author} {\bibfnamefont {J.~C.}\ \bibnamefont
  {Jim\'enez}}\ and\ \bibinfo {author} {\bibfnamefont {E.~S.}\ \bibnamefont
  {Fraga}},\ }\bibfield  {title} {\bibinfo {title} {{Radial oscillations of
  quark stars admixed with dark matter}},\ }\href@noop {} {\  (\bibinfo {year}
  {2021})},\ \Eprint {https://arxiv.org/abs/2111.00091} {arXiv:2111.00091
  [hep-ph]} \BibitemShut {NoStop}%
\bibitem [{\citenamefont {Brillante}\ and\ \citenamefont
  {Mishustin}(2014)}]{Brillante:2014lwa}%
  \BibitemOpen
  \bibfield  {author} {\bibinfo {author} {\bibfnamefont {A.}~\bibnamefont
  {Brillante}}\ and\ \bibinfo {author} {\bibfnamefont {I.~N.}\ \bibnamefont
  {Mishustin}},\ }\bibfield  {title} {\bibinfo {title} {{Radial oscillations of
  neutral and charged hybrid stars}},\ }\href
  {https://doi.org/10.1209/0295-5075/105/39001} {\bibfield  {journal} {\bibinfo
   {journal} {EPL}\ }\textbf {\bibinfo {volume} {105}},\ \bibinfo {pages}
  {39001} (\bibinfo {year} {2014})},\ \Eprint {https://arxiv.org/abs/1401.7915}
  {arXiv:1401.7915 [astro-ph.SR]} \BibitemShut {NoStop}%
\bibitem [{\citenamefont {Sagun}\ \emph {et~al.}(2020)\citenamefont {Sagun},
  \citenamefont {Panotopoulos},\ and\ \citenamefont {Lopes}}]{Sagun:2020qvc}%
  \BibitemOpen
  \bibfield  {author} {\bibinfo {author} {\bibfnamefont {V.}~\bibnamefont
  {Sagun}}, \bibinfo {author} {\bibfnamefont {G.}~\bibnamefont
  {Panotopoulos}},\ and\ \bibinfo {author} {\bibfnamefont {I.}~\bibnamefont
  {Lopes}},\ }\bibfield  {title} {\bibinfo {title} {{Asteroseismology: radial
  oscillations of neutron stars with realistic equation of state}},\ }\href
  {https://doi.org/10.1103/PhysRevD.101.063025} {\bibfield  {journal} {\bibinfo
   {journal} {Phys. Rev. D}\ }\textbf {\bibinfo {volume} {101}},\ \bibinfo
  {pages} {063025} (\bibinfo {year} {2020})},\ \Eprint
  {https://arxiv.org/abs/2002.12209} {arXiv:2002.12209 [astro-ph.HE]}
  \BibitemShut {NoStop}%
\bibitem [{\citenamefont {Sun}\ \emph {et~al.}(2021)\citenamefont {Sun},
  \citenamefont {Zheng}, \citenamefont {Chen}, \citenamefont {Burgio},\ and\
  \citenamefont {Schulze}}]{Sun:2021cez}%
  \BibitemOpen
  \bibfield  {author} {\bibinfo {author} {\bibfnamefont {T.-T.}\ \bibnamefont
  {Sun}}, \bibinfo {author} {\bibfnamefont {Z.-Y.}\ \bibnamefont {Zheng}},
  \bibinfo {author} {\bibfnamefont {H.}~\bibnamefont {Chen}}, \bibinfo {author}
  {\bibfnamefont {G.~F.}\ \bibnamefont {Burgio}},\ and\ \bibinfo {author}
  {\bibfnamefont {H.-J.}\ \bibnamefont {Schulze}},\ }\bibfield  {title}
  {\bibinfo {title} {{Equation of state and radial oscillations of neutron
  stars}},\ }\href {https://doi.org/10.1103/PhysRevD.103.103003} {\bibfield
  {journal} {\bibinfo  {journal} {Phys. Rev. D}\ }\textbf {\bibinfo {volume}
  {103}},\ \bibinfo {pages} {103003} (\bibinfo {year} {2021})},\ \Eprint
  {https://arxiv.org/abs/2101.07515} {arXiv:2101.07515 [nucl-th]} \BibitemShut
  {NoStop}%
\bibitem [{\citenamefont {Chandrasekhar}(1964)}]{Chandrasekhar:1964zz}%
  \BibitemOpen
  \bibfield  {author} {\bibinfo {author} {\bibfnamefont {S.}~\bibnamefont
  {Chandrasekhar}},\ }\bibfield  {title} {\bibinfo {title} {{The Dynamical
  Instability of Gaseous Masses Approaching the Schwarzschild Limit in General
  Relativity}},\ }\href {https://doi.org/10.1086/147938} {\bibfield  {journal}
  {\bibinfo  {journal} {Astrophys. J.}\ }\textbf {\bibinfo {volume} {140}},\
  \bibinfo {pages} {417} (\bibinfo {year} {1964})},\ \bibinfo {note} {[Erratum:
  Astrophys.J. 140, 1342 (1964)]}\BibitemShut {NoStop}%
\bibitem [{\citenamefont {Gleiser}(1988)}]{Gleiser:1988rq}%
  \BibitemOpen
  \bibfield  {author} {\bibinfo {author} {\bibfnamefont {M.}~\bibnamefont
  {Gleiser}},\ }\bibfield  {title} {\bibinfo {title} {{Stability of Boson
  Stars}},\ }\href {https://doi.org/10.1103/PhysRevD.38.2376} {\bibfield
  {journal} {\bibinfo  {journal} {Phys. Rev. D}\ }\textbf {\bibinfo {volume}
  {38}},\ \bibinfo {pages} {2376} (\bibinfo {year} {1988})},\ \bibinfo {note}
  {[Erratum: Phys.Rev.D 39, 1257 (1989)]}\BibitemShut {NoStop}%
\bibitem [{\citenamefont {Jetzer}(1989)}]{Jetzer:1988vr}%
  \BibitemOpen
  \bibfield  {author} {\bibinfo {author} {\bibfnamefont {P.}~\bibnamefont
  {Jetzer}},\ }\bibfield  {title} {\bibinfo {title} {{Dynamical Instability of
  Bosonic Stellar Configurations}},\ }\href
  {https://doi.org/10.1016/0550-3213(89)90038-2} {\bibfield  {journal}
  {\bibinfo  {journal} {Nucl. Phys. B}\ }\textbf {\bibinfo {volume} {316}},\
  \bibinfo {pages} {411} (\bibinfo {year} {1989})}\BibitemShut {NoStop}%
\bibitem [{\citenamefont {Gleiser}\ and\ \citenamefont
  {Watkins}(1989)}]{Gleiser:1988ih}%
  \BibitemOpen
  \bibfield  {author} {\bibinfo {author} {\bibfnamefont {M.}~\bibnamefont
  {Gleiser}}\ and\ \bibinfo {author} {\bibfnamefont {R.}~\bibnamefont
  {Watkins}},\ }\bibfield  {title} {\bibinfo {title} {{Gravitational Stability
  of Scalar Matter}},\ }\href {https://doi.org/10.1016/0550-3213(89)90627-5}
  {\bibfield  {journal} {\bibinfo  {journal} {Nucl. Phys. B}\ }\textbf
  {\bibinfo {volume} {319}},\ \bibinfo {pages} {733} (\bibinfo {year}
  {1989})}\BibitemShut {NoStop}%
\bibitem [{\citenamefont {Kain}(2021{\natexlab{c}})}]{Kain:2021rmk}%
  \BibitemOpen
  \bibfield  {author} {\bibinfo {author} {\bibfnamefont {B.}~\bibnamefont
  {Kain}},\ }\bibfield  {title} {\bibinfo {title} {{Boson stars and their
  radial oscillations}},\ }\href {https://doi.org/10.1103/PhysRevD.103.123003}
  {\bibfield  {journal} {\bibinfo  {journal} {Phys. Rev. D}\ }\textbf {\bibinfo
  {volume} {103}},\ \bibinfo {pages} {123003} (\bibinfo {year}
  {2021}{\natexlab{c}})},\ \Eprint {https://arxiv.org/abs/2106.01740}
  {arXiv:2106.01740 [gr-qc]} \BibitemShut {NoStop}%
\bibitem [{\citenamefont {Comer}\ \emph {et~al.}(1999)\citenamefont {Comer},
  \citenamefont {Langlois},\ and\ \citenamefont {Lin}}]{Comer:1999rs}%
  \BibitemOpen
  \bibfield  {author} {\bibinfo {author} {\bibfnamefont {G.}~\bibnamefont
  {Comer}}, \bibinfo {author} {\bibfnamefont {D.}~\bibnamefont {Langlois}},\
  and\ \bibinfo {author} {\bibfnamefont {L.~M.}\ \bibnamefont {Lin}},\
  }\bibfield  {title} {\bibinfo {title} {{Quasinormal modes of general
  relativistic superfluid neutron stars}},\ }\href
  {https://doi.org/10.1103/PhysRevD.60.104025} {\bibfield  {journal} {\bibinfo
  {journal} {Phys. Rev. D}\ }\textbf {\bibinfo {volume} {60}},\ \bibinfo
  {pages} {104025} (\bibinfo {year} {1999})},\ \Eprint
  {https://arxiv.org/abs/gr-qc/9908040} {arXiv:gr-qc/9908040} \BibitemShut
  {NoStop}%
\bibitem [{\citenamefont {Goldman}\ and\ \citenamefont
  {Nussinov}(1989)}]{Goldman:1989nd}%
  \BibitemOpen
  \bibfield  {author} {\bibinfo {author} {\bibfnamefont {I.}~\bibnamefont
  {Goldman}}\ and\ \bibinfo {author} {\bibfnamefont {S.}~\bibnamefont
  {Nussinov}},\ }\bibfield  {title} {\bibinfo {title} {{Weakly Interacting
  Massive Particles and Neutron Stars}},\ }\href
  {https://doi.org/10.1103/PhysRevD.40.3221} {\bibfield  {journal} {\bibinfo
  {journal} {Phys. Rev. D}\ }\textbf {\bibinfo {volume} {40}},\ \bibinfo
  {pages} {3221} (\bibinfo {year} {1989})}\BibitemShut {NoStop}%
\bibitem [{\citenamefont {Kouvaris}(2008)}]{Kouvaris:2007ay}%
  \BibitemOpen
  \bibfield  {author} {\bibinfo {author} {\bibfnamefont {C.}~\bibnamefont
  {Kouvaris}},\ }\bibfield  {title} {\bibinfo {title} {{WIMP Annihilation and
  Cooling of Neutron Stars}},\ }\href
  {https://doi.org/10.1103/PhysRevD.77.023006} {\bibfield  {journal} {\bibinfo
  {journal} {Phys. Rev. D}\ }\textbf {\bibinfo {volume} {77}},\ \bibinfo
  {pages} {023006} (\bibinfo {year} {2008})},\ \Eprint
  {https://arxiv.org/abs/0708.2362} {arXiv:0708.2362 [astro-ph]} \BibitemShut
  {NoStop}%
\bibitem [{\citenamefont {Bertone}\ and\ \citenamefont
  {Fairbairn}(2008)}]{Bertone:2007ae}%
  \BibitemOpen
  \bibfield  {author} {\bibinfo {author} {\bibfnamefont {G.}~\bibnamefont
  {Bertone}}\ and\ \bibinfo {author} {\bibfnamefont {M.}~\bibnamefont
  {Fairbairn}},\ }\bibfield  {title} {\bibinfo {title} {{Compact Stars as Dark
  Matter Probes}},\ }\href {https://doi.org/10.1103/PhysRevD.77.043515}
  {\bibfield  {journal} {\bibinfo  {journal} {Phys. Rev. D}\ }\textbf {\bibinfo
  {volume} {77}},\ \bibinfo {pages} {043515} (\bibinfo {year} {2008})},\
  \Eprint {https://arxiv.org/abs/0709.1485} {arXiv:0709.1485 [astro-ph]}
  \BibitemShut {NoStop}%
\bibitem [{\citenamefont {de~Lavallaz}\ and\ \citenamefont
  {Fairbairn}(2010)}]{deLavallaz:2010wp}%
  \BibitemOpen
  \bibfield  {author} {\bibinfo {author} {\bibfnamefont {A.}~\bibnamefont
  {de~Lavallaz}}\ and\ \bibinfo {author} {\bibfnamefont {M.}~\bibnamefont
  {Fairbairn}},\ }\bibfield  {title} {\bibinfo {title} {{Neutron Stars as Dark
  Matter Probes}},\ }\href {https://doi.org/10.1103/PhysRevD.81.123521}
  {\bibfield  {journal} {\bibinfo  {journal} {Phys. Rev. D}\ }\textbf {\bibinfo
  {volume} {81}},\ \bibinfo {pages} {123521} (\bibinfo {year} {2010})},\
  \Eprint {https://arxiv.org/abs/1004.0629} {arXiv:1004.0629 [astro-ph.GA]}
  \BibitemShut {NoStop}%
\bibitem [{\citenamefont {Kouvaris}\ and\ \citenamefont
  {Tinyakov}(2010)}]{Kouvaris:2010vv}%
  \BibitemOpen
  \bibfield  {author} {\bibinfo {author} {\bibfnamefont {C.}~\bibnamefont
  {Kouvaris}}\ and\ \bibinfo {author} {\bibfnamefont {P.}~\bibnamefont
  {Tinyakov}},\ }\bibfield  {title} {\bibinfo {title} {{Can Neutron stars
  constrain Dark Matter?}},\ }\href
  {https://doi.org/10.1103/PhysRevD.82.063531} {\bibfield  {journal} {\bibinfo
  {journal} {Phys. Rev. D}\ }\textbf {\bibinfo {volume} {82}},\ \bibinfo
  {pages} {063531} (\bibinfo {year} {2010})},\ \Eprint
  {https://arxiv.org/abs/1004.0586} {arXiv:1004.0586 [astro-ph.GA]}
  \BibitemShut {NoStop}%
\bibitem [{\citenamefont {Cerme\~no}\ \emph {et~al.}(2017)\citenamefont
  {Cerme\~no}, \citenamefont {P\'erez-Garc\'\i{}a},\ and\ \citenamefont
  {Silk}}]{Cermeno:2017xwb}%
  \BibitemOpen
  \bibfield  {author} {\bibinfo {author} {\bibfnamefont {M.}~\bibnamefont
  {Cerme\~no}}, \bibinfo {author} {\bibfnamefont {M.~A.}\ \bibnamefont
  {P\'erez-Garc\'\i{}a}},\ and\ \bibinfo {author} {\bibfnamefont
  {J.}~\bibnamefont {Silk}},\ }\bibfield  {title} {\bibinfo {title} {{Fermionic
  Light Dark Matter Particles and the New Physics of Neutron Stars}},\ }\href
  {https://doi.org/10.1017/pasa.2017.38} {\bibfield  {journal} {\bibinfo
  {journal} {Publ. Astron. Soc. Austral.}\ }\textbf {\bibinfo {volume} {34}},\
  \bibinfo {pages} {e043} (\bibinfo {year} {2017})},\ \Eprint
  {https://arxiv.org/abs/1710.06866} {arXiv:1710.06866 [astro-ph.HE]}
  \BibitemShut {NoStop}%
\bibitem [{\citenamefont {Seidel}\ and\ \citenamefont
  {Suen}(1994)}]{Seidel:1993zk}%
  \BibitemOpen
  \bibfield  {author} {\bibinfo {author} {\bibfnamefont {E.}~\bibnamefont
  {Seidel}}\ and\ \bibinfo {author} {\bibfnamefont {W.-M.}\ \bibnamefont
  {Suen}},\ }\bibfield  {title} {\bibinfo {title} {{Formation of solitonic
  stars through gravitational cooling}},\ }\href
  {https://doi.org/10.1103/PhysRevLett.72.2516} {\bibfield  {journal} {\bibinfo
   {journal} {Phys. Rev. Lett.}\ }\textbf {\bibinfo {volume} {72}},\ \bibinfo
  {pages} {2516} (\bibinfo {year} {1994})},\ \Eprint
  {https://arxiv.org/abs/gr-qc/9309015} {arXiv:gr-qc/9309015} \BibitemShut
  {NoStop}%
\bibitem [{\citenamefont {Di~Giovanni}\ \emph {et~al.}(2018)\citenamefont
  {Di~Giovanni}, \citenamefont {Sanchis-Gual}, \citenamefont {Herdeiro},\ and\
  \citenamefont {Font}}]{DiGiovanni:2018bvo}%
  \BibitemOpen
  \bibfield  {author} {\bibinfo {author} {\bibfnamefont {F.}~\bibnamefont
  {Di~Giovanni}}, \bibinfo {author} {\bibfnamefont {N.}~\bibnamefont
  {Sanchis-Gual}}, \bibinfo {author} {\bibfnamefont {C.~A.~R.}\ \bibnamefont
  {Herdeiro}},\ and\ \bibinfo {author} {\bibfnamefont {J.~A.}\ \bibnamefont
  {Font}},\ }\bibfield  {title} {\bibinfo {title} {{Dynamical formation of
  Proca stars and quasistationary solitonic objects}},\ }\href
  {https://doi.org/10.1103/PhysRevD.98.064044} {\bibfield  {journal} {\bibinfo
  {journal} {Phys. Rev. D}\ }\textbf {\bibinfo {volume} {98}},\ \bibinfo
  {pages} {064044} (\bibinfo {year} {2018})},\ \Eprint
  {https://arxiv.org/abs/1803.04802} {arXiv:1803.04802 [gr-qc]} \BibitemShut
  {NoStop}%
\bibitem [{\citenamefont {Romero}\ \emph {et~al.}(1996)\citenamefont {Romero},
  \citenamefont {Ibanez}, \citenamefont {Marti},\ and\ \citenamefont
  {Miralles}}]{Romero:1995cn}%
  \BibitemOpen
  \bibfield  {author} {\bibinfo {author} {\bibfnamefont {J.~V.}\ \bibnamefont
  {Romero}}, \bibinfo {author} {\bibfnamefont {J.~M.}\ \bibnamefont {Ibanez}},
  \bibinfo {author} {\bibfnamefont {J.~M.}\ \bibnamefont {Marti}},\ and\
  \bibinfo {author} {\bibfnamefont {J.~A.}\ \bibnamefont {Miralles}},\
  }\bibfield  {title} {\bibinfo {title} {{A new spherically symmetric general
  relativistic hydrodynamical code}},\ }\href {https://doi.org/10.1086/177198}
  {\bibfield  {journal} {\bibinfo  {journal} {Astrophys. J.}\ }\textbf
  {\bibinfo {volume} {462}},\ \bibinfo {pages} {839} (\bibinfo {year}
  {1996})},\ \Eprint {https://arxiv.org/abs/astro-ph/9509121}
  {arXiv:astro-ph/9509121} \BibitemShut {NoStop}%
\bibitem [{\citenamefont {Rezzolla}\ and\ \citenamefont
  {Zanotti}(2013)}]{RezzollaBook}%
  \BibitemOpen
  \bibfield  {author} {\bibinfo {author} {\bibfnamefont {L.}~\bibnamefont
  {Rezzolla}}\ and\ \bibinfo {author} {\bibfnamefont {O.}~\bibnamefont
  {Zanotti}},\ }\href@noop {} {\emph {\bibinfo {title} {{Relativistic
  Hydrodynamics}}}}\ (\bibinfo  {publisher} {Oxford},\ \bibinfo {address}
  {Oxford, UK},\ \bibinfo {year} {2013})\BibitemShut {NoStop}%
\bibitem [{\citenamefont {Neilsen}\ and\ \citenamefont
  {Choptuik}(2000)}]{Neilsen:1999we}%
  \BibitemOpen
  \bibfield  {author} {\bibinfo {author} {\bibfnamefont {D.~W.}\ \bibnamefont
  {Neilsen}}\ and\ \bibinfo {author} {\bibfnamefont {M.~W.}\ \bibnamefont
  {Choptuik}},\ }\bibfield  {title} {\bibinfo {title} {{Ultrarelativistic fluid
  dynamics}},\ }\href {https://doi.org/10.1088/0264-9381/17/4/302} {\bibfield
  {journal} {\bibinfo  {journal} {Class. Quant. Grav.}\ }\textbf {\bibinfo
  {volume} {17}},\ \bibinfo {pages} {733} (\bibinfo {year} {2000})},\ \Eprint
  {https://arxiv.org/abs/gr-qc/9904052} {arXiv:gr-qc/9904052} \BibitemShut
  {NoStop}%
\bibitem [{\citenamefont {Guzm\'an}\ \emph {et~al.}(2012)\citenamefont
  {Guzm\'an}, \citenamefont {Lora-Clavijo},\ and\ \citenamefont
  {Morales}}]{Guzman:2012rn}%
  \BibitemOpen
  \bibfield  {author} {\bibinfo {author} {\bibfnamefont {F.~S.}\ \bibnamefont
  {Guzm\'an}}, \bibinfo {author} {\bibfnamefont {F.~D.}\ \bibnamefont
  {Lora-Clavijo}},\ and\ \bibinfo {author} {\bibfnamefont {M.~D.}\ \bibnamefont
  {Morales}},\ }\bibfield  {title} {\bibinfo {title} {{Revisiting spherically
  symmetric relativistic hydrodynamics}},\ }\href@noop {} {\bibfield  {journal}
  {\bibinfo  {journal} {Rev. Mex. Fis. E}\ }\textbf {\bibinfo {volume} {58}},\
  \bibinfo {pages} {84} (\bibinfo {year} {2012})},\ \Eprint
  {https://arxiv.org/abs/1212.1421} {arXiv:1212.1421 [gr-qc]} \BibitemShut
  {NoStop}%
\bibitem [{\citenamefont {Shu}\ and\ \citenamefont {Osher}(1988)}]{SHU1988439}%
  \BibitemOpen
  \bibfield  {author} {\bibinfo {author} {\bibfnamefont {C.-W.}\ \bibnamefont
  {Shu}}\ and\ \bibinfo {author} {\bibfnamefont {S.}~\bibnamefont {Osher}},\
  }\bibfield  {title} {\bibinfo {title} {Efficient implementation of
  essentially non-oscillatory shock-capturing schemes},\ }\href
  {https://doi.org/https://doi.org/10.1016/0021-9991(88)90177-5} {\bibfield
  {journal} {\bibinfo  {journal} {Journal of Computational Physics}\ }\textbf
  {\bibinfo {volume} {77}},\ \bibinfo {pages} {439} (\bibinfo {year}
  {1988})}\BibitemShut {NoStop}%
\bibitem [{\citenamefont {Okun}(2007)}]{Okun:2006eb}%
  \BibitemOpen
  \bibfield  {author} {\bibinfo {author} {\bibfnamefont {L.}~\bibnamefont
  {Okun}},\ }\bibfield  {title} {\bibinfo {title} {{Mirror particles and mirror
  matter: 50 years of speculations and search}},\ }\href
  {https://doi.org/10.1070/PU2007v050n04ABEH006227} {\bibfield  {journal}
  {\bibinfo  {journal} {Phys. Usp.}\ }\textbf {\bibinfo {volume} {50}},\
  \bibinfo {pages} {380} (\bibinfo {year} {2007})},\ \Eprint
  {https://arxiv.org/abs/hep-ph/0606202} {arXiv:hep-ph/0606202} \BibitemShut
  {NoStop}%
\bibitem [{\citenamefont {Hippert}\ \emph {et~al.}(2021)\citenamefont
  {Hippert}, \citenamefont {Setford}, \citenamefont {Tan}, \citenamefont
  {Curtin}, \citenamefont {Noronha-Hostler},\ and\ \citenamefont
  {Yunes}}]{Hippert:2021fch}%
  \BibitemOpen
  \bibfield  {author} {\bibinfo {author} {\bibfnamefont {M.}~\bibnamefont
  {Hippert}}, \bibinfo {author} {\bibfnamefont {J.}~\bibnamefont {Setford}},
  \bibinfo {author} {\bibfnamefont {H.}~\bibnamefont {Tan}}, \bibinfo {author}
  {\bibfnamefont {D.}~\bibnamefont {Curtin}}, \bibinfo {author} {\bibfnamefont
  {J.}~\bibnamefont {Noronha-Hostler}},\ and\ \bibinfo {author} {\bibfnamefont
  {N.}~\bibnamefont {Yunes}},\ }\bibfield  {title} {\bibinfo {title} {{Mirror
  Neutron Stars}},\ }\href@noop {} {\  (\bibinfo {year} {2021})},\ \Eprint
  {https://arxiv.org/abs/2103.01965} {arXiv:2103.01965 [astro-ph.HE]}
  \BibitemShut {NoStop}%
\bibitem [{\citenamefont {Glass}\ and\ \citenamefont
  {Lindblom}(1983)}]{Glass1983}%
  \BibitemOpen
  \bibfield  {author} {\bibinfo {author} {\bibfnamefont {E.~N.}\ \bibnamefont
  {Glass}}\ and\ \bibinfo {author} {\bibfnamefont {L.}~\bibnamefont
  {Lindblom}},\ }\bibfield  {title} {\bibinfo {title} {{The Radial Oscillations
  of Neutron Stars}},\ }\href@noop {} {\bibfield  {journal} {\bibinfo
  {journal} {Astrophys. J. Supp}\ }\textbf {\bibinfo {volume} {53}},\ \bibinfo
  {pages} {93} (\bibinfo {year} {1983})}\BibitemShut {NoStop}%
\bibitem [{\citenamefont {Gondek}\ \emph {et~al.}(1997)\citenamefont {Gondek},
  \citenamefont {Haensel},\ and\ \citenamefont {Zdunik}}]{Gondek:1997fd}%
  \BibitemOpen
  \bibfield  {author} {\bibinfo {author} {\bibfnamefont {D.}~\bibnamefont
  {Gondek}}, \bibinfo {author} {\bibfnamefont {P.}~\bibnamefont {Haensel}},\
  and\ \bibinfo {author} {\bibfnamefont {J.}~\bibnamefont {Zdunik}},\
  }\bibfield  {title} {\bibinfo {title} {{Radial pulsations and stability of
  protoneutron stars}},\ }\href@noop {} {\bibfield  {journal} {\bibinfo
  {journal} {Astron. Astrophys.}\ }\textbf {\bibinfo {volume} {325}},\ \bibinfo
  {pages} {217} (\bibinfo {year} {1997})},\ \Eprint
  {https://arxiv.org/abs/astro-ph/9705157} {arXiv:astro-ph/9705157}
  \BibitemShut {NoStop}%
\bibitem [{\citenamefont {Kokkotas}\ and\ \citenamefont
  {Ruoff}(2001)}]{Kokkotas:2000up}%
  \BibitemOpen
  \bibfield  {author} {\bibinfo {author} {\bibfnamefont {K.}~\bibnamefont
  {Kokkotas}}\ and\ \bibinfo {author} {\bibfnamefont {J.}~\bibnamefont
  {Ruoff}},\ }\bibfield  {title} {\bibinfo {title} {{Radial oscillations of
  relativistic stars}},\ }\href {https://doi.org/10.1051/0004-6361:20000216}
  {\bibfield  {journal} {\bibinfo  {journal} {Astron. Astrophys.}\ }\textbf
  {\bibinfo {volume} {366}},\ \bibinfo {pages} {565} (\bibinfo {year}
  {2001})},\ \Eprint {https://arxiv.org/abs/gr-qc/0011093}
  {arXiv:gr-qc/0011093} \BibitemShut {NoStop}%
\bibitem [{\citenamefont {Gabler}\ \emph {et~al.}(2009)\citenamefont {Gabler},
  \citenamefont {Sperhake},\ and\ \citenamefont {Andersson}}]{Gabler:2009yt}%
  \BibitemOpen
  \bibfield  {author} {\bibinfo {author} {\bibfnamefont {M.}~\bibnamefont
  {Gabler}}, \bibinfo {author} {\bibfnamefont {U.}~\bibnamefont {Sperhake}},\
  and\ \bibinfo {author} {\bibfnamefont {N.}~\bibnamefont {Andersson}},\
  }\bibfield  {title} {\bibinfo {title} {{Nonlinear radial oscillations of
  neutron stars}},\ }\href {https://doi.org/10.1103/PhysRevD.80.064012}
  {\bibfield  {journal} {\bibinfo  {journal} {Phys. Rev. D}\ }\textbf {\bibinfo
  {volume} {80}},\ \bibinfo {pages} {064012} (\bibinfo {year} {2009})},\
  \Eprint {https://arxiv.org/abs/0906.3088} {arXiv:0906.3088 [gr-qc]}
  \BibitemShut {NoStop}%
\bibitem [{\citenamefont {Jetzer}(1990)}]{Jetzer:1990xa}%
  \BibitemOpen
  \bibfield  {author} {\bibinfo {author} {\bibfnamefont {P.}~\bibnamefont
  {Jetzer}},\ }\bibfield  {title} {\bibinfo {title} {{Stability of Combined
  Boson - Fermion Stars}},\ }\href
  {https://doi.org/10.1016/0370-2693(90)90952-3} {\bibfield  {journal}
  {\bibinfo  {journal} {Phys. Lett. B}\ }\textbf {\bibinfo {volume} {243}},\
  \bibinfo {pages} {36} (\bibinfo {year} {1990})}\BibitemShut {NoStop}%
\end{thebibliography}

%

\end{document}